# Polygrammar: Grammar for Digital Polymer Representation and Generation

*Minghao Guo,[1,2] Wan Shou,[1] Liane Makatura,[1] Timothy Erps,[1] Michael Foshey,[1] Wojciech Matusik[1]**

[1]Computer Science and Artificial Intelligence Lab, Massachusetts Institute of Technology, Cambridge, Massachusetts, United States
[2]CUHK Multimedia Lab, The Chinese University of Hong Kong, Sha Tin, Hong Kong

*Corresponding author. Email: wojciech@csail.mit.edu

Polymers are widely-studied materials with diverse properties and applications determined by different molecular structures. It is essential to represent these structures clearly and explore the full space of achievable chemical designs. However, existing approaches are unable to offer comprehensive design models for polymers because of their inherent scale and structural complexity. Here, we present a parametric, context-sensitive grammar designed specifically for the representation and generation of polymers. As a demonstrative example, we implement our grammar for polyurethanes. Using our symbolic hypergraph representation and 14 simple production rules, our PolyGrammar is able to represent and generate all valid polyurethane structures. We also present an algorithm to translate any polyurethane structure from the popular SMILES string format into our PolyGrammar representation. We test the representative power of PolyGrammar by translating a dataset of over 600 polyurethane samples collected from literature. Furthermore, we show that PolyGrammar can be easily extended to the other copolymers and homopolymers such as polyacrylates. By offering a complete, explicit representation scheme and an explainable generative model with validity guarantees, our PolyGrammar takes an important step toward a more comprehensive and practical system for polymer discovery and exploration. As the first bridge between formal languages and chemistry, PolyGrammar also serves as a critical blueprint to inform the design of similar grammars for other chemistries, including organic and inorganic molecules.

## 1. Introduction

Polymers are important materials with diverse structure variations, and applications. To facilitate customized applications and deepen the fundamental understanding, it is extremely beneficial to characterize, enumerate, and explore the full space of achievable polymer structures. Such a large (ideally exhaustive) collection of polymers would be particularly powerful in conjunction with machine learning and numerical simulation techniques, as a way to facilitate complicated tasks like human-guided molecular exploration[1-12], property prediction[13-17], and retro-synthesis[18,19]. Computational approaches based on chemical representations and generated data[20-27] have also tremendously reduced the time, cost, and resources spent on physical synthesis in the chemistry lab[28-31].

Ideally, a chemical design model would include three components: (1) a well-defined *representation* capable of capturing known structures, (2) a *generative model* capable of enumerating all structures in a given class, and (3) an *inverse modeling procedure* capable of translating known molecular structures into the representation. For a given class of molecules, an ideal chemical design model should satisfy the following five criteria: i. **Complete:** representation is able to encode all possible structures in the given class. ii. **Explicit:** representation directly specifies the molecular structure. iii. **Valid:** every generated output is a physically valid chemical structure in the given class. iv. **Explainable:** the generation process is understandable to the user. v. **Invertible:** the inverse procedure can translate molecular structures into the given representation. However, designing a chemical model that meets all these criteria is challenging, especially for structurally complex molecules. Most existing approaches are limited to small, simple chemical structures[32-36]. Even with this limited scope, the design is labor intensive: the representation language is typically developed first, then extended for generation and inverse modeling. In particular, there have been many systems for molecular line notations[32,33] and fragment-level description[34,35], which were then used as the basis for generative and inverse schemes[5-8].

Yet, a comprehensive chemical design model for large polymers remains elusive due to the polymers' inherent complexity. We present a detailed account for each property, including polymer-specific challenges and the performance of existing methods (see Table 1). Some of the most popular methods like SMILES and BigSMILES are only partial design models, as they define a representation but not a generative model. In this case, we assume the simplest generative model for our comparison: randomly chosen strings of permissible symbols. Other chemical design models like auto-encoders (AE)[5-8,36] have a direct mapping to our framework: the learned latent space is the representation, the decoder is the generative model, and the encoder is the inverse model. After exploring the state of the art for all five properties, we give an overview of our proposed approach.

**Table 1. Comparison with related chemical design models.** Since SMILES and BigSMILES only explicitly provide a representation, we assume the simplest generative modeling scheme: randomly choosing strings of permissible symbols. Our PolyGrammar is the only approach that satisfies all five properties.

| | **Representation** | | **Generative Modeling** | | **Inverse Modeling** |
|---|---|---|---|---|---|
| **Methods** | **Complete** | **Explicit** | **Valid** | **Explainable** | **Translation from SMILES** |
| **SMILES**[28] | √ | √ | × | √ | √ |
| **BigSMILES**[33] | √ | × | × | × | √ |
| **Auto-Encoders**[5-8, 36] | × | × | × | × | √ |
| **PolyGrammar** | √ | √ | √ | √ | √ |

**Complete.** Polymers are intrinsically stochastic molecules constructed from some distribution of chemical sub-units. Thus, given a particular set of reactants, the synthesized polymers are not unique; rather, there is wide variation in the resulting structures. For example, consider the polyurethanes synthesized by a 1:1 ratio of two distinct components: methylene

bis(phenyl isocyanate) (MDI) and poly(oxytetramethylene) diol (PTMO). Consider one possible outcome of chain length 6, where *chain length* is defined as the sum of MDI and PTMO units. Disregarding more nuanced chemical restrictions (which are beyond the scope of this paper), any arrangement of the 3 MDI and 3 PTMO units is equally valid. Thus, for a chain of length 20, the component permutations can result in more than $\binom{20}{10} \approx 10^5$ possible structures. This vast set of structures makes it challenging to design a complete and concise polymer representation. Some existing line notations[28-34] including SMILES[28] (designed for general molecules) and BigSMILES[33] (specifically designed for large polymers) are complete representations, since they can convert any given polymer structure instance into the form of strings. However, schemes relying on auto-encoders (AE) are not guaranteed to satisfy this property since the learned representation spaces (numeric vectors called *latent variables*) may exclude polymer structures that do not exist in training data.

**Explicit.** The properties of a polymeric material are largely determined by the structure of the polymer itself, including the identity and arrangement of its constituent monomers[37-39]. Thus, it is useful to have an *explicit* representation for polymers, in which specific structural information is directly expressed and easily understood. This is challenging because a polymer must be understood on many scales, including the overarching structure of repeated units, and the individual molecular and atomic sub-units that comprise them.

Low-level representations like SMILES are able to depict explicit polymeric structures, but the strings are typically hard to parse due to their length. For example, the canonical SMILES representation for the polyurethane chain of length 30 (5 repetitions of the 6-length chain described above) requires more than 600 characters. By contrast, most representations designed for large polymers[32-34] are so high-level that they are unable to provide explicit information about the complete polymer structure. For example, BigSMILES can express the constituent monomers and the bonding descriptions between them, but it cannot specify the detailed arrangement of the polymer's components. As for the AE, the latent variable is an implicit representation and it is impractical to understand the polymer structures merely from the numeric vector.

**Valid.** Generative models that build on a well-defined representation scheme are highly coveted[40], particularly for their ability to efficiently build large corpora of example structures. However, the result is only useful if the examples generated by the model are guaranteed to be chemically valid. This is challenging to enforce for polymers, as there are many hard chemical constraints (e.g., valency conditions) and other restrictions to account for. The likelihood of violating these constraints increases as the target molecules get larger.

Machine learning techniques including support vector machines (SVM)[41], recurrent neural networks (RNN)[1-4], generative adversarial networks (GAN)[9-12], and AE have been used as generative models for molecules. However, these methods often produce outputs that are chemically invalid, even when limited to small molecules. It is even more challenging for these methods to generate valid polymers, due to the large number of generation steps required to realize such large molecules. Although several recent efforts based on AE[35,36] and reinforcement learning (RL)[42,43] have been proposed to produce valid polymers, it is not clear how well they generalize – i.e., the AE may be unable to ensure validity when generating polymers that significantly deviate from the training data. Non-learning methods also struggle to enforce validity, particularly with simple probabilistic generative models – e.g., randomly choosing SMILES/BigSMILES strings. Even with additional considerations for line notation syntax and additional semantic constraints, these probabilistic generation schemes can produce invalid line notations[44].

**Explainable.** To ensure confidence in the results of the generative model, the generation process itself must be fully transparent and understandable to chemists. This property is not necessarily more challenging for large polymers (compared to small molecules), but it is much more critical to facilitate understanding of the resulting polymer structure. Interpretable generation processes also aid the exploration of possible polymer variations.

AE and other deep learning based generative models[1-4,9,10,45] produce structures based on implicit latent variables. These models are effectively black-box functions that cannot be easily interpreted. By contrast, the generative model of SMILES can be interpreted since each generated symbol has an explainable meaning: it either indicates the type of the generated atom, or the bonding relationship. The generative model based on BigSMILES is not explainable since it cannot show the detailed arrangement of constituent monomers.

**Invertible.** When designing a new chemical design model, it is critical to ensure compatibility with existing notations. In particular, it should be possible (via an *inverse modelling procedure*) to translate any final representation from an existing scheme into the proposed representation. This inverse procedure should yield the same process and final representation as if the structure were created via the integrated generative model. This is critical for two reasons: (i) it makes existing knowledge accessible in the new representation, and (ii) it confirms the representative power of the new chemical design model.

To judge invertibility for polymer models, we consider translation from one of the most popular molecule notations: SMILES. As shown in Table 1, invertibility is already an important feature common to many existing methods. For example, the encoder of a chemical AE takes a SMILES string as input, then outputs the corresponding latent variable. BigSMILES is built directly upon SMILES so it can easily covert SMILES strings of polymers into the BigSMILES representation. When building our own representation, we also consider "invertibility" with respect to the SMILES format. However, in principle, it is possible to design inverse procedures that translate from other existing representations schemes as well.

**Our Approach.** In this paper, we propose a new chemical design model for polymers that respects all five of the ideal properties discussed above. We introduce *PolyGrammar*, a parametric context-sensitive grammar for polymers. In formal language theory, a *grammar* describes how to build strings from a language's *alphabet* following a set of *production rules*. PolyGrammar represents the chain structure as a hypergraph. In particular, each polymer chain is represented as a string of symbols, each of which refers to a particular molecular fragment of the original chain. This symbolic hypergraph representation supports explicit descriptions for infinite amount of diversely structured polymer chains by changing the form of symbolic strings.

Based on this representation, we establish a set of production rules that can effectively generate chemically valid symbolic strings. The recursive nature of grammar production makes it possible to generate any polymer in our given class using only a simple set of production rules. In particular, it is possible for PolyGrammar to enumerate *all* valid polymers structures within a given class.

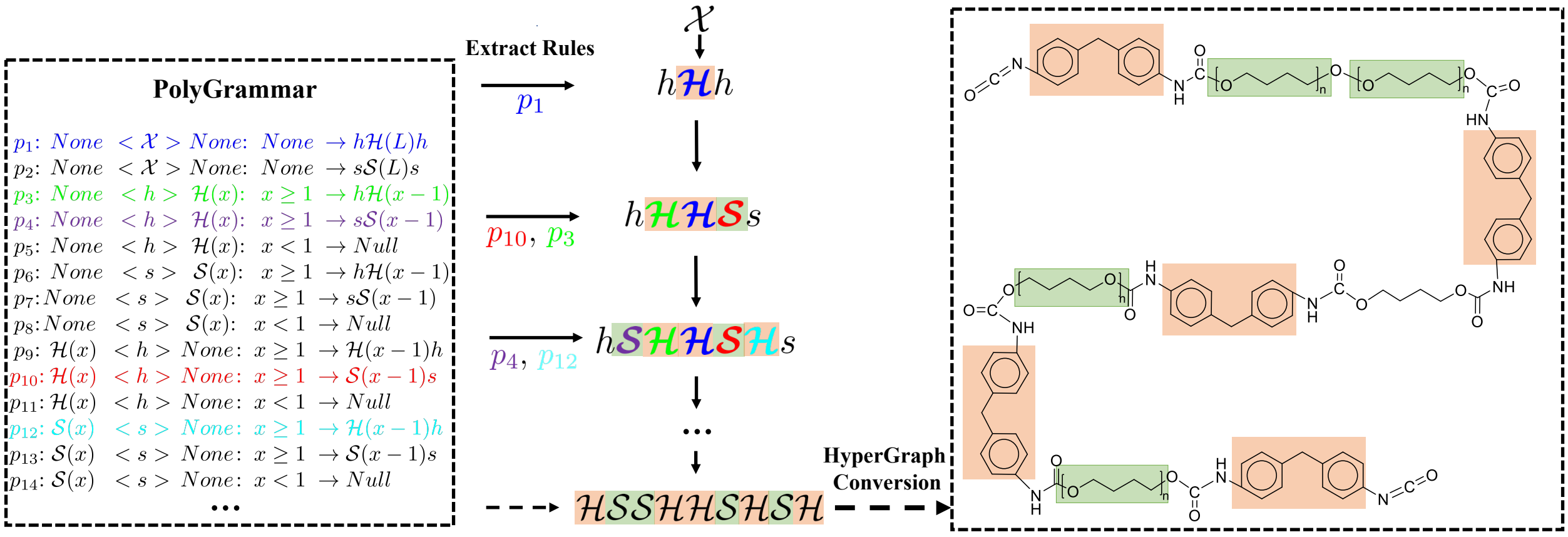

**Figure 1.** Schematic of our chemistry design model, *PolyGrammar*, which represents molecular chain structure as a string of symbols (center). PolyGrammar consists of a set of production rules $\{p_i | i = 1, \ldots, 14\}$ (left). The generation process starts from an initial symbol $\mathcal{X}$. At each iteration, each non-terminal symbol ($h$, $s$ or $\mathcal{X}$) in the current string is replaced by the successor of a production rule whose predecessor matches the symbol. The generation process concludes when the string does not contain any non-terminal symbols. The resulting symbol string (center) is then translated to a polymer chain (right) by hypergraph conversion.

As a demonstrative example, we focus on a particular class of polymers: *polyurethanes*. We choose polyurethanes due to their wide-ranging applications, including antistatic coating[46], foams[47], elastomers[48], and drug delivery for cancer therapy[49]. Consider generating a polyurethane of chain length of 20, using 1 polyol type (e.g., PTMO) and 1 isocynate type (e.g., MDI). Under these assumptions (which are representative of the average polyurethane chain[50]), PolyGrammar can generate more than $2 \times 10^6$ distinct polyurethane chains using only 14 production rules. Moreover, we show that PolyGrammar can be easily extended to the other types of polymers, including both copolymers and homopolymers. We further propose an inverse modeling algorithm that translates a polymer's SMILES string into the sequence of production rules used to generate it. More than 600 polyurethanes collected from literature are validated by this inverse model, demonstrating the representative power of PolyGrammar. Schematic of our PolyGrammar is shown in Figure 1.

## 2. Hypergraph-based Symbolic Representation

In this section, we introduce the hypergraph representation of polyurethane structures and describe how to use symbolic strings to represent polyurethane chains.

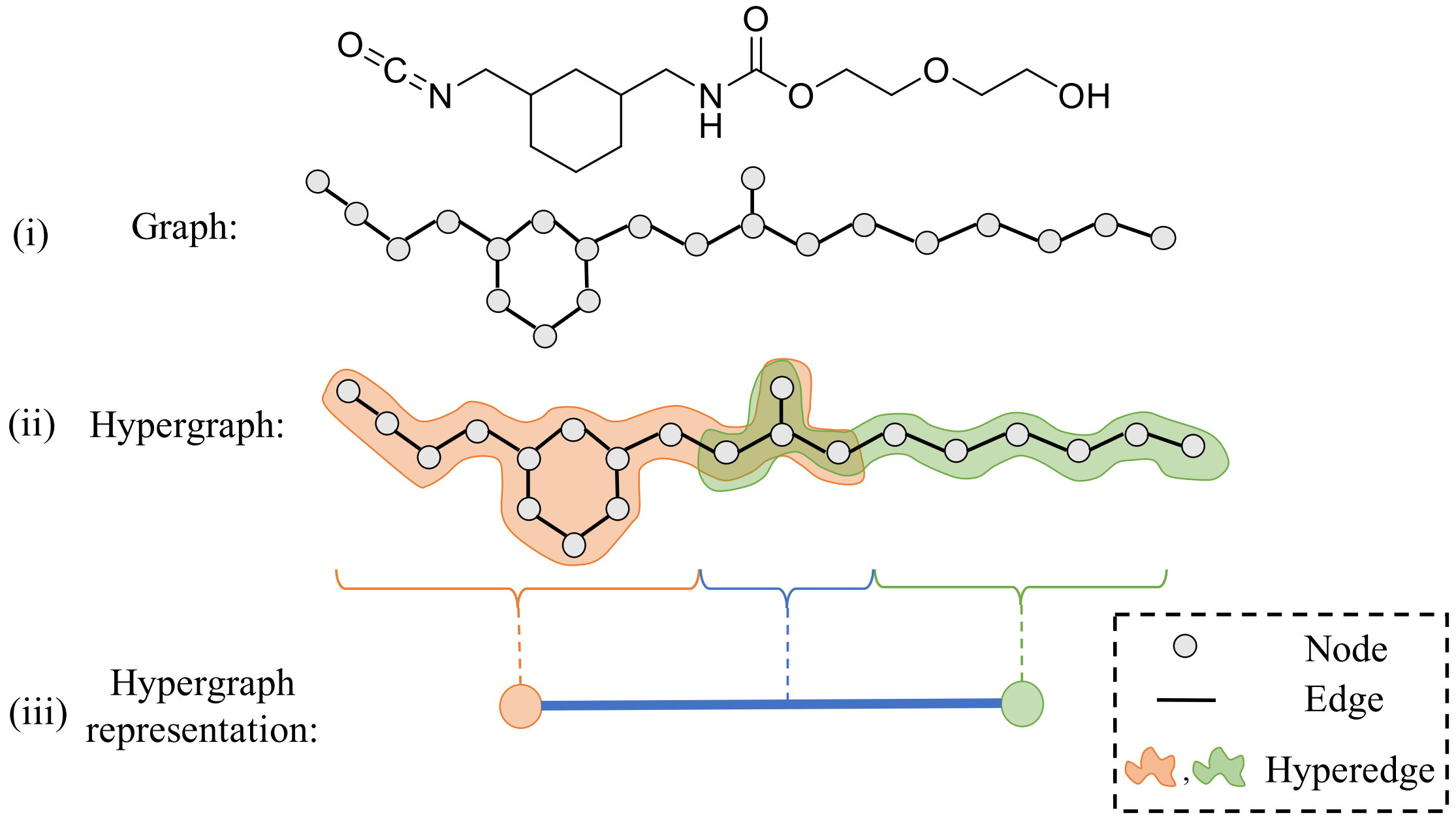

**Figure 2.** The structure produced by reacted by two monomers (1,3-bis(isocyanatomethyl)cyclohexane and diethylene glycol). The standard graph representation (i) uses 21 nodes and 21 edges, but the hypergraph (ii) only requires two hyperedges. Each hyperedge corresponds to the nodes of a given monomer. Both hyperedges have the urethane group in common. We use the line graph (iii) to visualize the hypergraph representation in the remaining figures of the paper for convenience.

### 2.1. Polymers as Hypergraphs

It is a common practice[7,12,51,52] to regard the structural formula of a molecule as an ordinary graph, where atoms are nodes, bonds are edges, and edges connect exactly two nodes. For polyurethanes, ordinary graph depictions would require prohibitively many nodes and edges. To address this, we employ a generalized graph called a *hypergraph*[53], which allows individual edges to join more than one node. Any edge that connects a subset of the nodes in the hypergraph is called a *hyperedge*. Consider the product of two monomers (1,3 bis(isocyanatomethyl)cyclohexane and diethylene glycol) as shown in Figure 2(i). Originally, the graph required 21 nodes and 21 edges. However, if we construct each hyperedge by selecting the subset of nodes according to the monomer type, as shown in Figure 2(ii), the hypergraph for this molecule requires only 2 hyperedges. This dramatically reduces the representation cost for large polyurethane chains.

For increased convenience, we will visualize the hypergraph representations using the *line graph*[54] form shown in Figure 2(iii). In graph theory, the line graph refers to the duality of the original graph, where each edge in the original graph corresponds to a unique vertex of the line graph. With regards to the theory of hypergraph, the line graph contains one vertex for every hyperedge in the original hypergraph. Two vertices in the line graph are connected by a line if their corresponding hyperedges in the original hypergraph have a non-empty intersection. For the hypergraph in Figure 2(ii), since the urethane group is shared by two hyperedges in the hypergraph, the corresponding line graph can be visualized as two vertices connected by an edge. By collapsing the original nodes based on molecular identity, the line graph form provides a more concise visualization of a hypergraph.

Complete polyurethane structures can also be represented in this manner. The molecular fragments corresponding to the isocyanate and the polyol in the polyurethane chain are represented as hyperedges, which are visualized as vertices in the line graph. The urethane groups connecting hard segment (HS) with soft segment (SS) and the chain extenders connecting two diisocyanates are viewed as intersections between two hyperedges; thus, they are visualized as edges in the line graph. Two examples of hypergraph representations for polyurethane structures are shown in Figure 3.

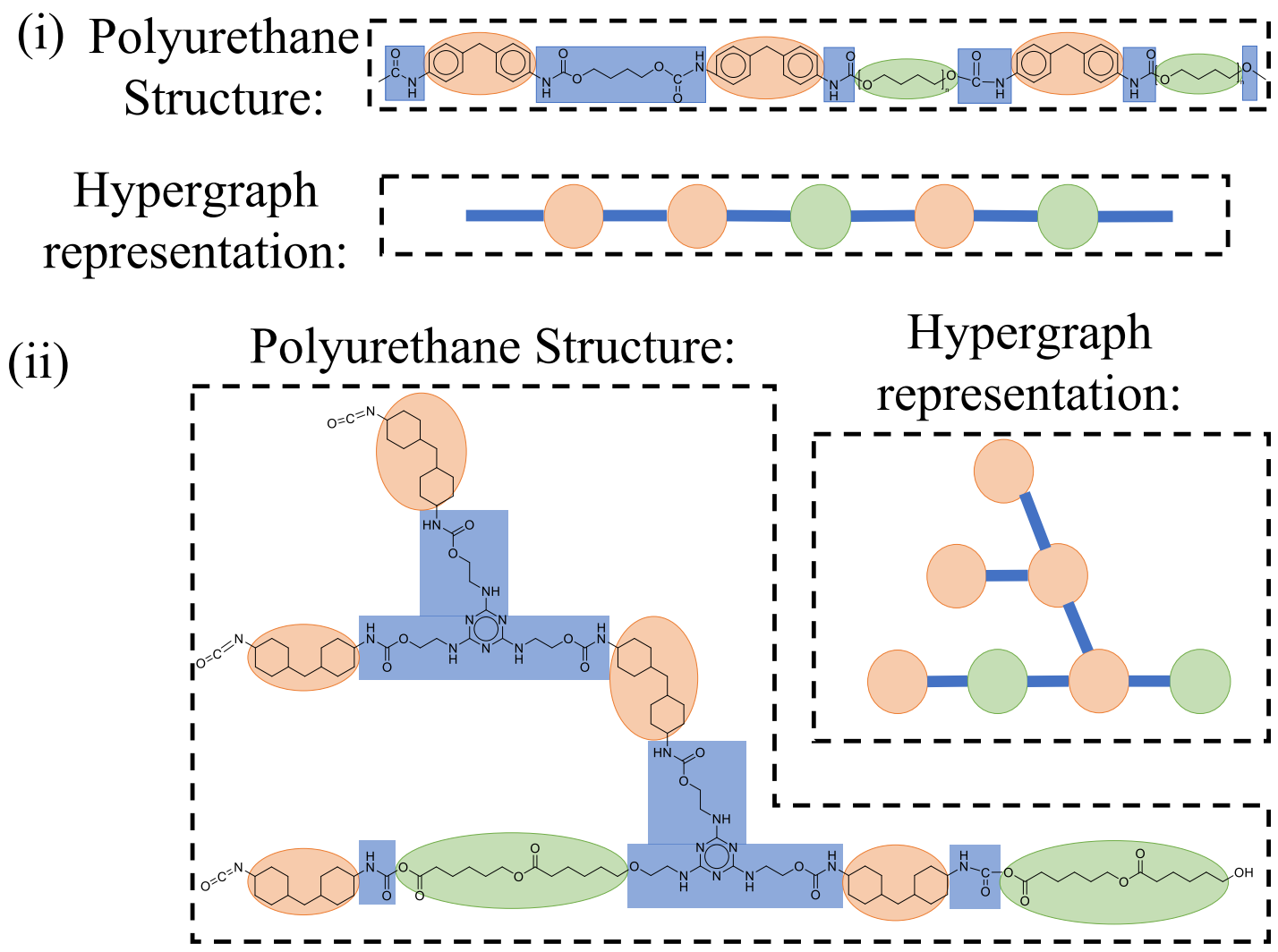

**Figure 3.** Examples of hypergraph representation. (i) Polyurethane chain synthesised by MDI, PTMO and 1,4-butanediol (BDO); (ii) Branched polyurethane chain synthesised by 4,4'-diisocyanato-methylenedicyclohexane (4,4'-HMDI), poly(caprolactone) diol (PCL) and triazine based polyhydric alcohol (3-THA).

### 2.2. Symbolic Representation

Given the hypergraph of a polyurethane chain, we construct a corresponding symbolic string for use in PolyGrammar. In the symbolic string, the hyperedges corresponding to the isocyanate (hard segment) are denoted with "$\mathcal{H}$" and those corresponding to the polyol (soft segment) are denoted as "$\mathcal{S}$". The chain extenders are omitted, since they can only exist between two adjacent $\mathcal{H}$ (or $\mathcal{S}$) symbols. For those polyurethanes containing multiple isocyanate or polyol types, we use subscripts $i = 1, 2, \ldots$ to distinguish different subtypes of certain hyperedge. For instance, if two different types of isocyanates are used[38], we use $\mathcal{H}_1$ and $\mathcal{H}_2$ to distinguish the hyperedges corresponding to each hard-segment type. These rules allow us to represent any polyurethane chain as a string of symbols. Examples are shown in Figure 4.

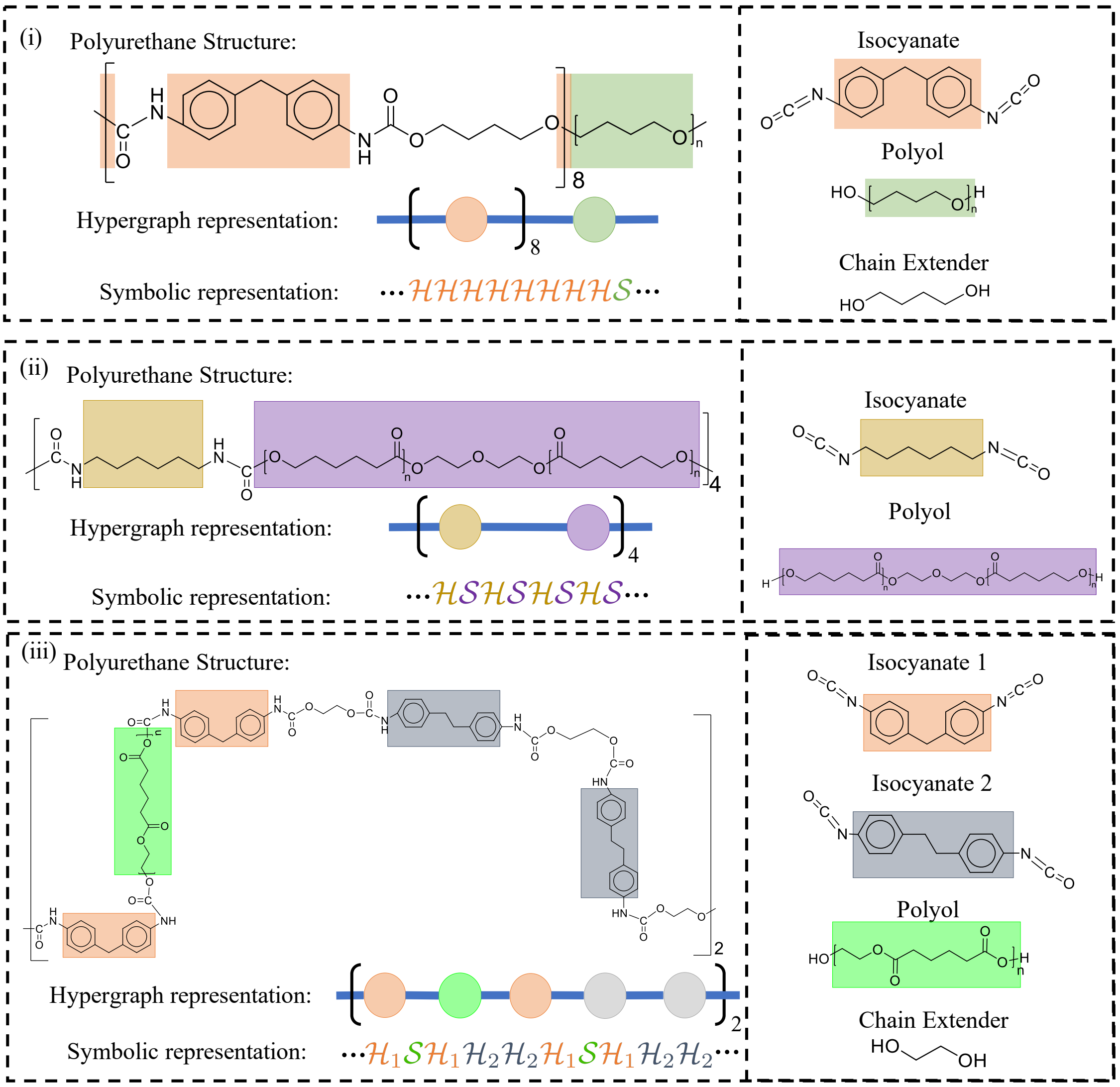

**Figure 4.** Symbolic representations for polyurethanes synthesized using: (i) MDI, PTMO and BDO; (ii) 1,6-diisocyanatohexane (HDI) and PCL; (iii) 4,4'-dibenzyl diisocyanate (DBDI), MDI, poly(ethylene adipate)diol (PEA) and ethylene glycol (EG). Note that (iii) includes multiple diisocyanates.

We emphasize that our symbolic representation is invertible, such that a symbolic string can be converted back to the corresponding chemical structure if the constituent isocyanate(s), polyol(s) and chain extender(s) are specified. We call this process *hypergraph conversion*. The invertibility of hypergraph representation ensures our PolyGrammar can simultaneously serve as a representation and a generative model for polyurethanes.

## 3. PolyGrammar

In this section, we first present the basic mechanism of grammar production using an illustrative example. Then, we introduce our parametric context-sensitive PolyGrammar comprehensively. Finally, we propose several advanced features based on our basic PolyGrammar for the representation of polyurethanes, which encourage the generation of more general structures.

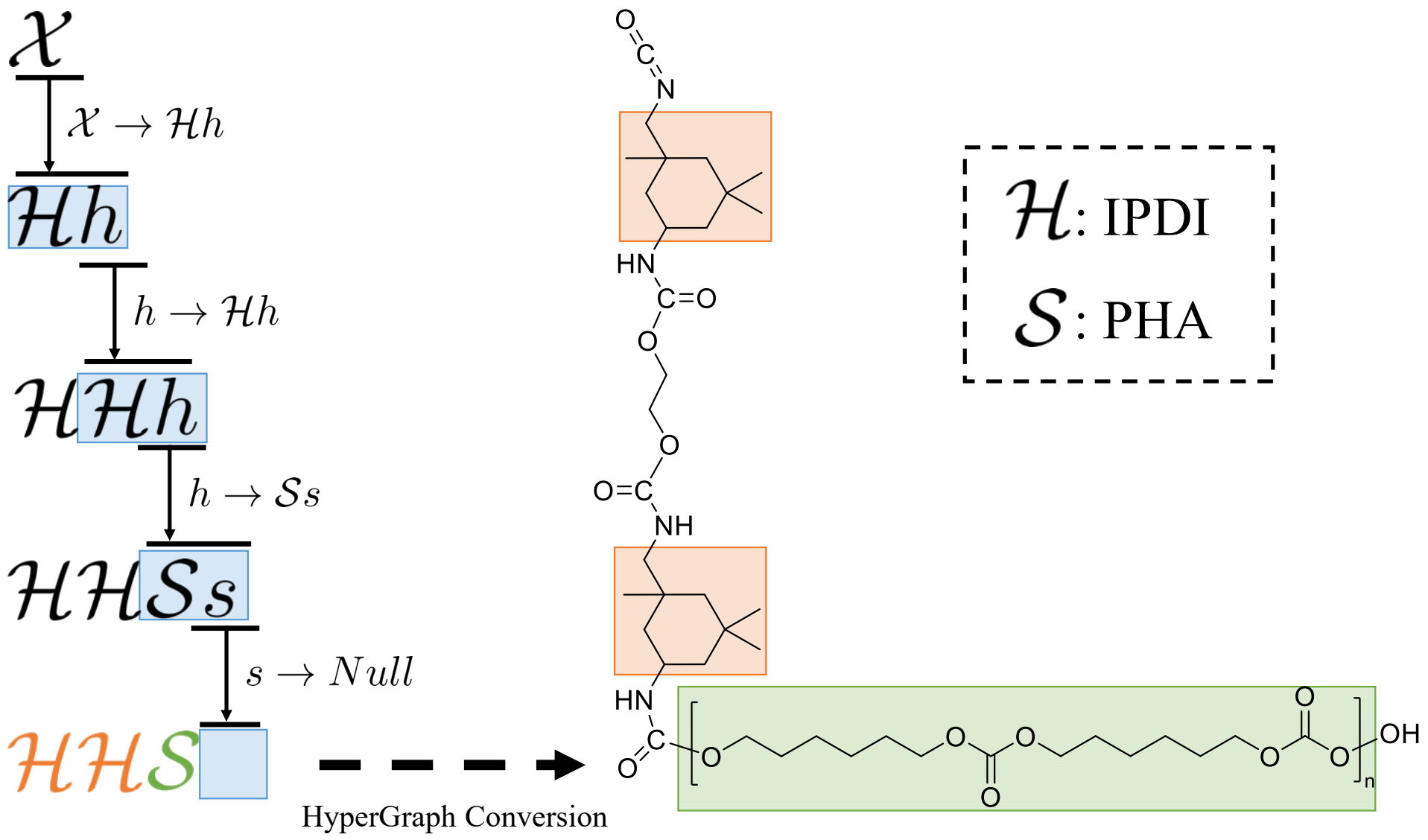

**Figure 5.** An illustrative example of grammar production. Starting from the initial symbol $\mathcal{X}$, we sequentially invoke four production rules from $P = \{\mathcal{X} \rightarrow \mathcal{H}h; h \rightarrow \mathcal{H}h; h \rightarrow \mathcal{S}s; s \rightarrow Null\}$. The process continues until all symbols in the string are terminal symbols. By specifying the constituent structures, i.e., isophorone diisocyanate (IPDI), polyhexamethylene carbonate glycol (PHA) and EG, the string of the symbols can be translated to the corresponding polyurethane chain via hypergraph conversion.

### 3.1. Basic PolyGrammar

In formal language theory, a grammar $G = (N, \Sigma, P)$ is used to describe a language, where $N$ is a set of non-terminal symbols, $\Sigma$ is a set of terminal symbols and $P$ is a set of production rules, each of which consists of a predecessor and a successor separated by a right arrow "→". In the language represented by the grammar $G$, each word is a finite-length string containing both terminal and non-terminal symbols. The non-terminal symbols in a word can be further replaced and expanded by invoking one production rule from $P$ at a step. In our PolyGrammar, the set of non-terminal symbols $N$ is $\{\mathcal{X}, h, s\}$ and the set of terminal symbols $\Sigma$ is $\{\mathcal{H}, \mathcal{S}\}$. Figure 5 shows an illustrative example to demonstrate the process for producing a string via the grammar $G$. This example uses four production rules: $P = \{\mathcal{X} \rightarrow \mathcal{H}h; h \rightarrow$ $\mathcal{H}h; h \rightarrow \mathcal{S}s; s \rightarrow Null\}$. Starting from the initial symbol $\mathcal{X}$, at each iteration, each non-terminal symbol in the current string is replaced with the successor of a production rule whose predecessor matches the symbol. The process continues until no non-terminal symbols exist in the string.

According to Chomsky's classification[55], the grammar used in this illustrative example is a Type-2 grammar, also called *context-free* grammar, where the predecessor of each production rule consists of only one single non-terminal symbol. Similar paradigms are also utilized in L-systems to model the morphology of organisms[56,57].

*3.1.1 Context-Sensitive Grammar*

The context-free grammar discussed above is insufficient to imitate the polyurethane generation process because the symbolic string can only expand along one direction; however, polyurethanes generally grow along two opposite directions to form chain structures. To address this, our PolyGrammar utilizes a *context-sensitive* grammar. In particular, our PolyGrammar is a Type-1 grammar, a more general form of Type-2 grammar[58], where the production rules also consider the context (i.e., the surrounding symbols) of the given non-terminal symbol within the string.

By considering the symbol contexts, the production rules of a context-sensitive grammar can explicitly depict the growing direction of the polyurethane chain. The production rules are as follows:

$$
\begin{aligned}
&p_1 : None < \mathcal{X} > None \rightarrow h\mathcal{H}h \\
&p_2 : None < \mathcal{X} > None \rightarrow s\mathcal{S}s \\
&p_3 : None < h > \mathcal{H} \rightarrow h\mathcal{H} \\
&p_4 : None < h > \mathcal{H} \rightarrow s\mathcal{S} \\
&p_5 : None < h > \mathcal{H} \rightarrow Null \\
&p_6 : None < s > \mathcal{S} \rightarrow h\mathcal{H} \\
&p_7 : None < s > \mathcal{S} \rightarrow s\mathcal{S}
\end{aligned}
$$

$$p_8 : None \ < s > \ \mathcal{S} \ \rightarrow Null$$

$$p_9 : \ \mathcal{H} \ < h > None \rightarrow \mathcal{H}h$$

$$p_{10} : \ \mathcal{H} \ < h > None \rightarrow \mathcal{S}s$$

$$p_{11} : \ \mathcal{H} \ < h > None \rightarrow Null$$

$$p_{12} : \ \mathcal{S} \ < s > None \rightarrow \mathcal{H}h$$

$$p_{13} : \ \mathcal{S} \ < s > None \rightarrow \mathcal{S}s$$

$$p_{14} : \ \mathcal{S} \ < s > None \rightarrow Null$$

In each production rule, the non-terminal symbol to be replaced is inside the angle brackets “< >” of the predecessor. The contexts are the symbols located at both sides of “< >” in the predecessor ($None$ indicates no constraints). The rule can only be deployed when both contexts of the symbol have been matched.

Each rule has an intuitive function. Rules $p_1$ and $p_2$ initialize the start symbol $\mathcal{X}$, while $p_5$, $p_8$, $p_{11}$ and $p_{14}$ terminate the growth. Rules $p_3$, $p_4$, $p_6$ and $p_7$ extend the string along the left direction, and $p_9$, $p_{10}$, $p_{12}$ and $p_{13}$ extend the string along the right direction. $p_3$ and $p_9$ indicate the reaction between two isocyanates, imitating the formation of the hard segment, while $p_7$ and $p_{13}$ indicate the reaction between two polyols, imitating the formation of the soft segment. Lastly, $p_4$, $p_6$, $p_{10}$ and $p_{12}$ imitate the formation of the urethane group.

Another important feature of the PolyGrammar is that there are multiple possible production rules to expand a given symbol. For instance, $p_3$, $p_4$ and $p_5$ share the same predecessor and expand the non-terminal symbol h along the left direction. There are many possible schemes for selecting among these options, including hand-tuned heuristics or manual intervention to guide the scheme toward particular results. For simplicity, we have implemented a uniformly random selection technique: at each iteration, we randomly sample one rule from all of the candidate rules that meet the contexts and apply it to the symbol. An example of the production process is illustrated in Figure 6.

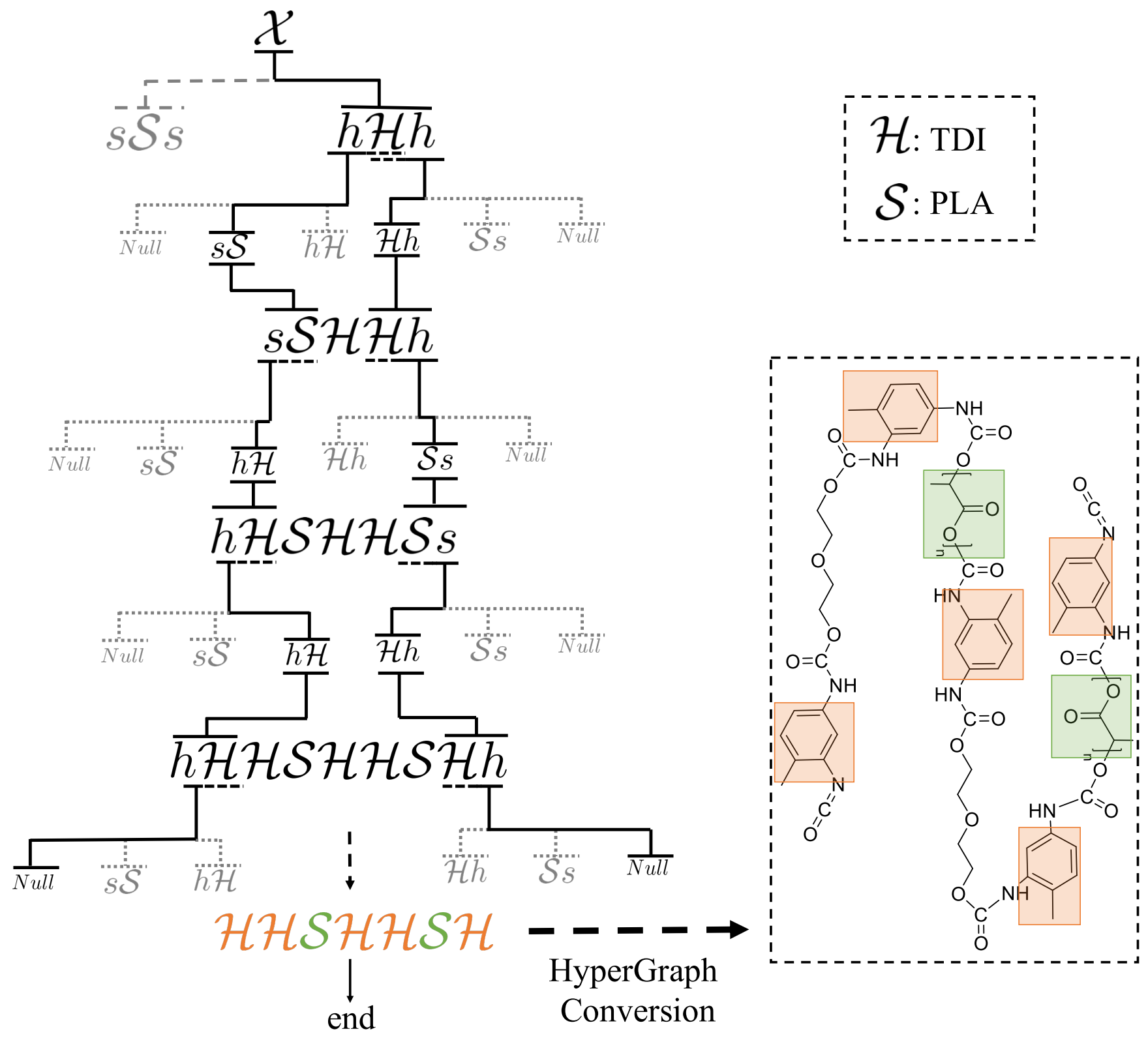

**Figure 6.** Example of context-sensitive grammar. At each production step, only the rules that match the non-terminal symbol's context are adopted. Hence, the production process can explicitly depict the growing direction of the polyurethane chain. If there are multiple candidate rules at a given step, selection can be done manually or randomly. The selected rule is then applied to the symbol to continue production.

### *3.1.2 Parametric Grammar*

Although the context-sensitive grammar makes it possible to generate a variety of polyurethane chain structures, its modeling power is still limited. One important problem is that the total chain length of the generated polyurethanes cannot be controlled. In practice, the chain length is an essential factor that influences the physical and chemical properties of the polyurethanes[50, 59]. It is non-trivial to control the chain length of each generated polyurethane merely using the grammar discussed above due to the stochastic production. In order to address this problem, we introduce a parameter $x$ associated with each terminal symbol in the grammar and augment our PolyGrammar as a *parametric* context-sensitive grammar. The proposed parametric grammar is illustrated as follows,

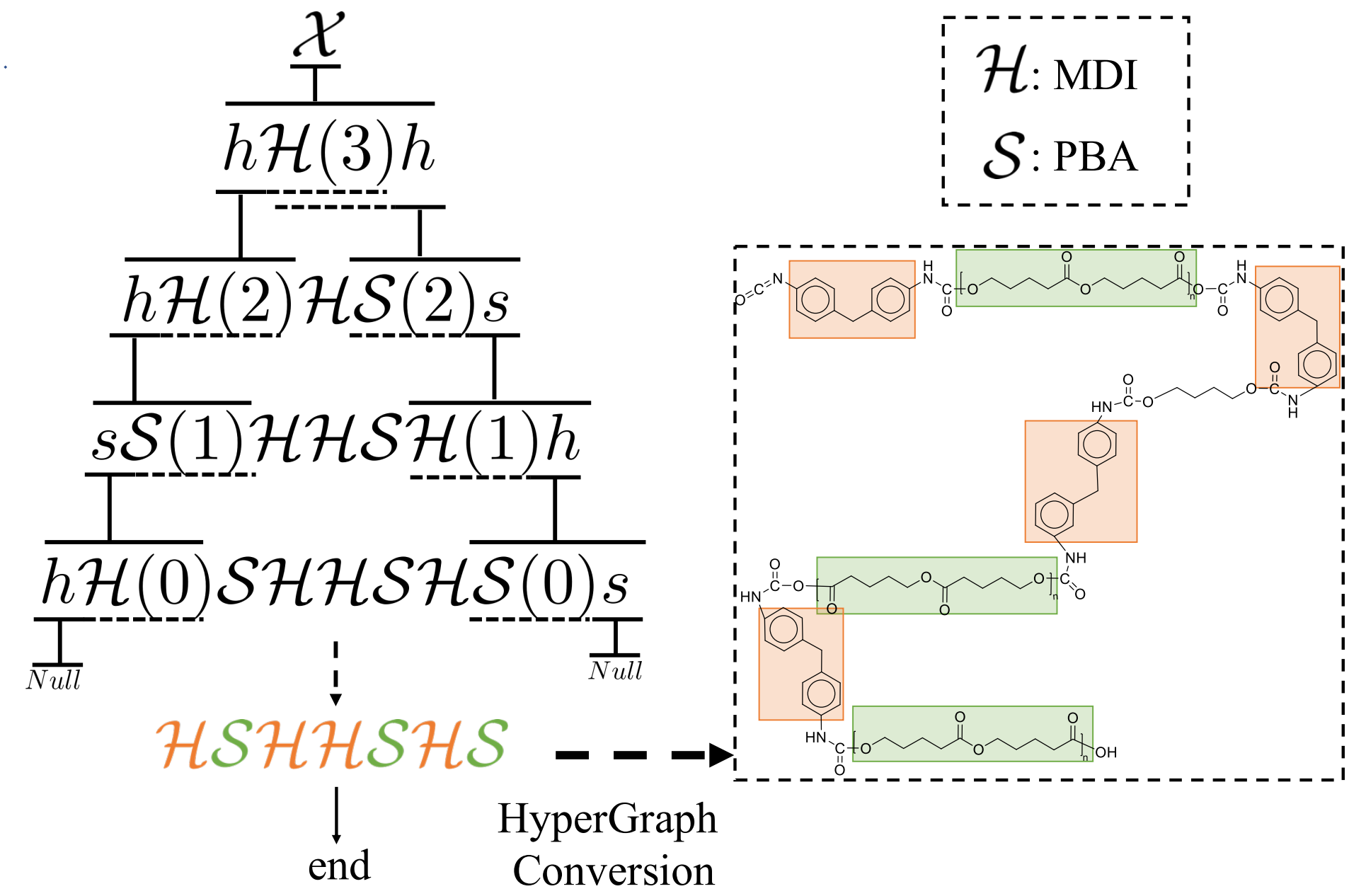

**Figure 7.** Example of parametric grammar. To control the length of generated polyurethane, we introduce parameters, denoted with parentheses "( )" after terminal symbols.

$$p_1 : None < \mathcal{X} > None: None \rightarrow h\mathcal{H}(L)h$$
$$p_2 : None < \mathcal{X} > None: None \rightarrow s\mathcal{S}(L)s$$
$$p_3 : None < h > \mathcal{H}(x): x \geq 1 \rightarrow h\mathcal{H}(x-1)$$
$$p_4 : None < h > \mathcal{H}(x): x \geq 1 \rightarrow s\mathcal{S}(x-1)$$
$$p_5 : None < h > \mathcal{H}(x): x < 1 \rightarrow Null$$
$$p_6 : None < s > \mathcal{S}(x): x \geq 1 \rightarrow h\mathcal{H}(x-1)$$
$$p_7 : None < s > \mathcal{S}(x): x \geq 1 \rightarrow s\mathcal{S}(x-1)$$
$$p_8 : None < s > \mathcal{S}(x): x < 1 \rightarrow Null$$
$$p_9 : \mathcal{H}(x) < h > None: x \geq 1 \rightarrow \mathcal{H}(x-1)h$$
$$p_{10}: \mathcal{H}(x) < h > None: x \geq 1 \rightarrow \mathcal{S}(x-1)s$$
$$p_{11}: \mathcal{H}(x) < h > None: x < 1 \rightarrow Null$$
$$p_{12}: \mathcal{S}(x) < s > None: x \geq 1 \rightarrow \mathcal{H}(x-1)h$$
$$p_{13}: \mathcal{S}(x) < s > None: x \geq 1 \rightarrow \mathcal{S}(x-1)s$$
$$p_{14}: \mathcal{S}(x) < s > None: x < 1 \rightarrow Null$$

The production rules now feature parameters, which are denoted with parentheses "( )" following terminal symbols. Furthermore, each production rule is augmented with a logical "condition" that determines whether the rule can be invoked or not (None indicates no constraints). By specifying L (the initial value of parameter x in production rules $p_1$ and $p_2$), the grammar can produce strings with length $2L + 1$, corresponding to polyurethane chains with length $2L + 1$. By varying the value of $L$, the chain length of generated polyurethanes can be controlled. An example of this production process is illustrated in Figure 7.

### 3.2. Advanced Features

*3.2.1 Extensions for Branched Polyurethanes*

So far, all of our polyurethanes have featured linear chain structures. However, it is possible for polyurethanes to have branched structures[60], as shown in Figure 3(ii). To generate branched polyurethanes, we augment the parametric context-sensitive grammar with several rules:

$$
\begin{aligned}
&p_{15}: None\ < h > \mathcal{H}(x): x \geq 1 \rightarrow h[h]\mathcal{H}(x-1) \\
&p_{16}: None\ < h > \mathcal{H}(x): x \geq 1 \rightarrow s[s]\mathcal{S}(x-1) \\
&p_{17}:\ \mathcal{H}(x) < h > None: x \geq 1 \rightarrow \mathcal{H}(x-1)[h]h \\
&p_{18}:\ \mathcal{S}(x)\ < s > None: x \geq 1 \rightarrow \mathcal{S}(x-1)[s]s
\end{aligned}
$$

A branch is delimited by the content inside a pair of square brackets "[ ]". The non-terminal symbols inside the square brackets can also be further expanded using the rules of the basic PolyGrammar. In the final string, all the terminal symbols inside a pair of square-brackets together form a sub-branch attached to the backbone. The above illustrated rules can generate polyurethane chains that have up to 2 branches at each bifurcation. The number of branches at each bifurcation can also exceed 2 by adding more square-bracket pairs attached to the non-terminal symbols. Examples are available in Supporting Information.

*3.2.2 Global Controllable Parameters*

We have already discussed the use of parameters for controlling the chain length of the generated polyurethanes. However, it is still difficult for our baseline parametric grammar to achieve more advanced controllable parameters such as the ratio of hard segment to soft segment. This is because the context-sensitive grammar only captures "local" information about the chain during the generation process, as the view of each production rule is limited to the context immediately surrounding the predecessor symbol. When it comes to global constraints, such as specific ratios of hard versus soft segments, the generative model needs to be aware of the relevant information (number of hard segments, chain length) over the whole chain. It is non-trivial to handle these constraints with the basic PolyGrammar discussed in previous sections.

To address this issue, we introduce an additional symbol "$\mathcal{M}$" which serves as a message that can collect global information about the chain. The message is propagated back and forth between the left and right ends of the string. The propagation is achieved by switching the message's position with the adjacent symbol's one at a time; this continues along a certain direction until the message gets to the string end. At each position swap, the message updates its parameters to collect the information required for the control setting. When the message reaches the end of the string, the outcome of the production rule is influenced by the information contained in the message. The message is then reset and begins to propagate along the opposite direction, encoding information about the entirety of the structure, continuing the above process. Since the production rules are only applied at the end of the chain, this mechanism ensures that the string generation adheres to all parameter-controlled constraints. Multiple constraints can be considered simultaneously by adding more parameters to the message symbol. The full set of the production rules and an illustration of the message passing mechanism are shown in Supporting Information.

**4. PolyGrammar as a Generative Model**

Generative models are critical for the efficient, thorough exploration of possible polymer structures. These models are also particularly powerful in conjunction with machine learning algorithms, in order to address complicated problems like human-guided exploration and property prediction. In this section, we discuss how our parametric context-sensitive PolyGrammar can serve as a generative model.

The generation process of PolyGrammar begins with a simple string that contains the initial symbol $\mathcal{X}$. On each step, we traverse the symbols in the current string and find the position of all the non-terminal symbols. For each non-terminal symbol, we identify a candidate set of production rules. Each candidate production rule must meet the following conditions: 1) the context in the predecessor clause matches the context of the current symbol in the string, and 2) the parameters of this symbol's context meet the logical condition of the production rule. If there are several candidate rules to expand a given symbol, a single rule is selected according to the desired scheme (random sampling scheme, manual intervention, etc.). We apply the selected production rule to the appropriate non-terminal symbol, and repeat this process until no non-terminal symbols remain in the string. Once the final string is produced, we convert it into an explicit polyurethane hypergraph by replacing the symbols in the string with the chemical structures (e.g., MDI and PTMO) corresponding to each hyperedge. This yields a valid, explicit polyurethane chain, as desired. These structures can be further converted to other forms of representation such as SMILES.

Using our generative model, it is possible to enumerate all valid polyurethane structures in a target class (e.g., length 20 with 1 type of polyol and 1 type of isocyanate). In particular, any distinct sequence of production rules on the start symbol yields a distinct string, which in turn represents a unique polyurethane chain. Since the production rules encode all permissible local configurations of the constituent molecules, it follows that our grammar is able to generate any valid polyurethane.

To emphasize the volume of achievable molecules, we also quantitatively analyze the diversity of generated chains for our PolyGrammar. Given a chain length parameter $L$ and the number of isocyanate and polyol types ($N_H$ and $N_S$, respectively), the basic PolyGrammar (with 14 production rules) allows the generation of a total number of

$$N = \sum_{i=0}^{2L+1} \frac{(2L+1)!}{i!\,(2L+1-i)!} N_H^i N_S^{2L+1-i}$$

polyurethane chains with different structures. With $L = 10, N_H = 1, N_S = 1$, which are representative of an average polyurethane chain[50], $N$ is more than $2 \times 10^6$. This demonstrates the powerful capacity of our PolyGrammar. Several polyurethane chains generated using PolyGrammar are shown in Figure 8. More examples can be found in Supporting Information.

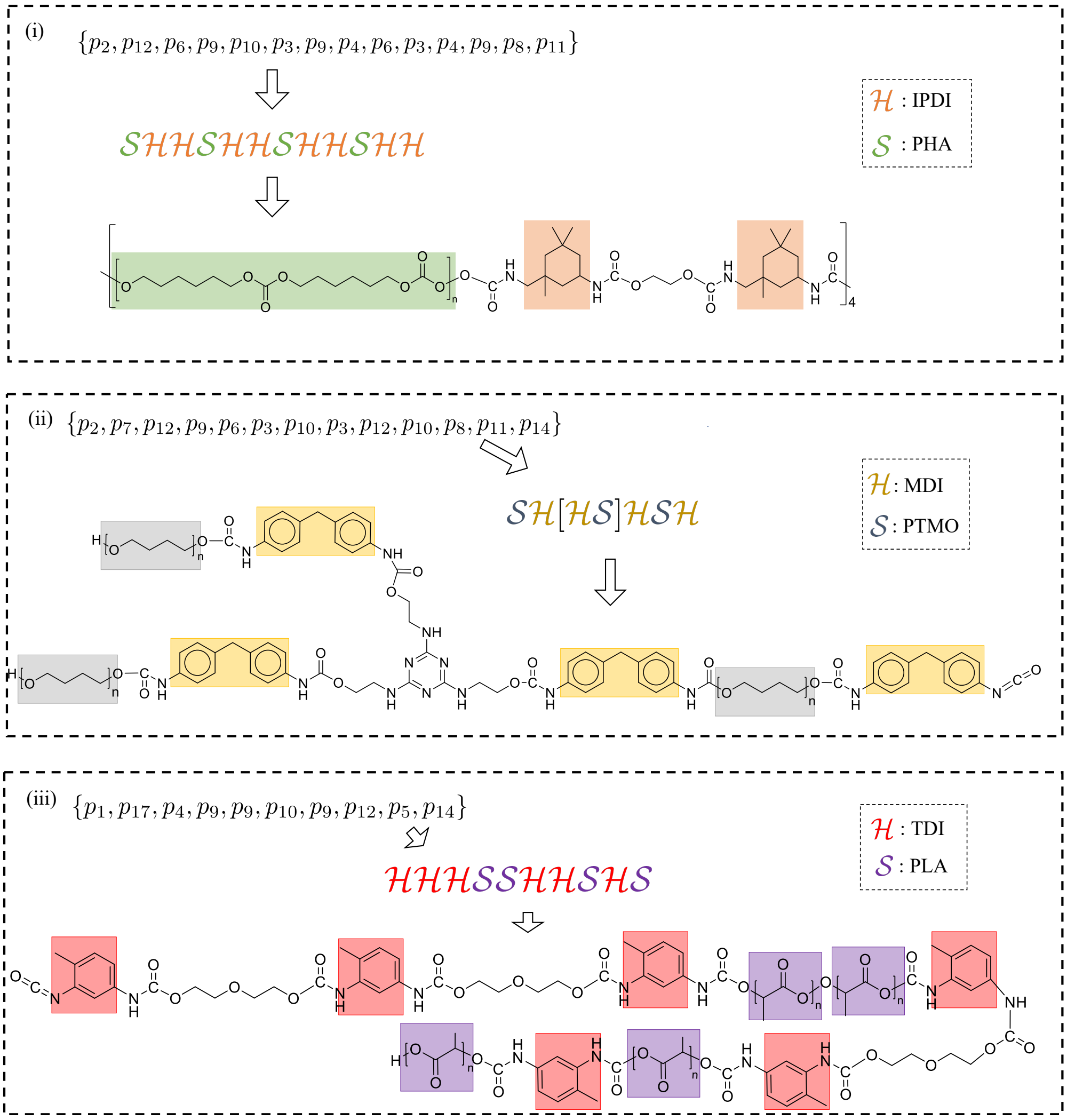

**Figure 8.** Examples of polyurethane chains generated using PolyGrammar. (i) Ordered chain with isophorone diisocyanate (IPDI), polyhexamethylene (PHA) and EG; (ii) Branched chain with MDI, PTMO and 3-THA; (iii) Unordered chain with Toluene diisocyanate (TDI), PLA and diethylene glycol (DEG).

## 5. Translation from SMILES

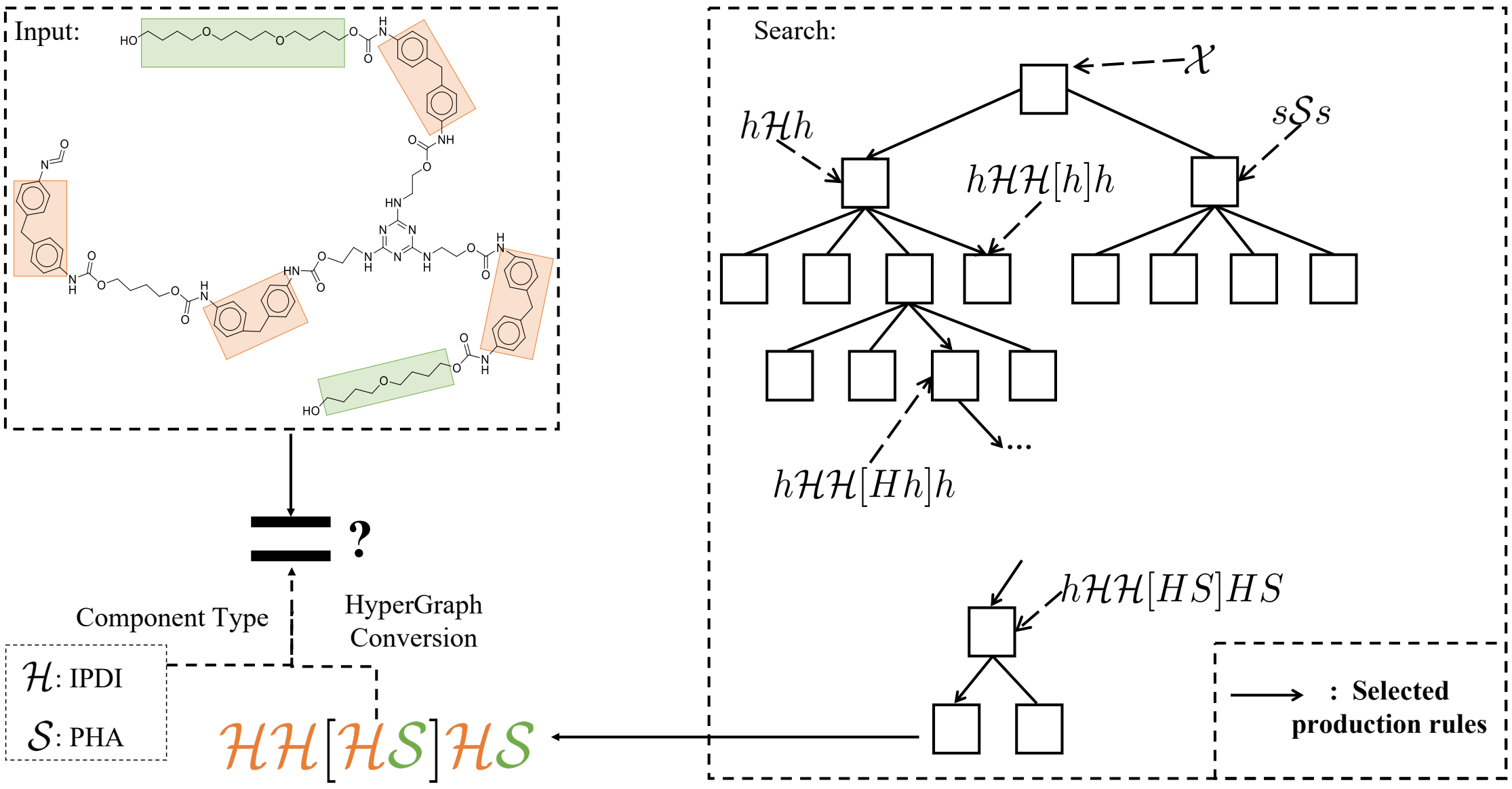

**Figure 9.** Schematic for translating a polyurethane from a SMILES string into our PolyGrammar representation, which also reveals the complete sequence of rules required for its generation. The pipeline can be regarded as a search process. Starting from the initial symbol, we iteratively select and invoke production rules until all symbols in the string are terminal symbols. Then given the component types, we convert the symbolic string into a polyurethane structure by hypergraph conversion and compare it with the input structure. The total process repeats until the search structure matches with the input structure.

To complete our chemical design model, we also develop an inverse model capable of translating a SMILES string into the corresponding sequence of PolyGrammar production rules. The overall pipeline of translation from SMILES can be regarded as a search process, as shown in Figure 9. Starting from the initial symbol, we iteratively select and invoke production rules until all symbols in the string are terminal symbols. Once we have a complete string and the specific component types, we use hypergraph conversion to convert the symbolic string into a polyurethane structure. We then compare our result with the input structure; if they do not match, we restart our search from scratch. The process repeats until our structure matches the original input.

Specifically, our inverse model proceeds as follows. Given the SMILES string of the polyurethane chain, we break it into multiple molecular fragments by disconnecting all of the urethane groups, $-NHCO-O-$. Then we exhaustively enumerate each molecular fragment and perform a string matching algorithm (KMP matching[63]) to identify the type of it: an isocyanate, a polyol or a chain extender. During the enumeration, we also record the connectivity between each fragment. Based on the types and the connectivity of the fragments, we can obtain a hypergraph representation of the original SMILES string. The final step is to convert the hypergraph into the sequence of the production rules of PolyGrammar. We traverse the hypergraph using the breadth-first search (BFS) algorithm, which explores all of the neighbouring hyperedges at the present depth before moving on to the nodes at the next depth level. BFS starts at the tree root, which is an arbitrary hyperedge of the hypergraph. Each step of the exploration returns a tuple of two hyperedges, which is then matched with a specific production rule in the PolyGrammar. Hence, the sequence of the production rules can be obtained once the entire hypergraph has been explored. The pipeline of this algorithm is illustrated in Figure 10 and the corresponding pseudo-code is in Supporting Information.

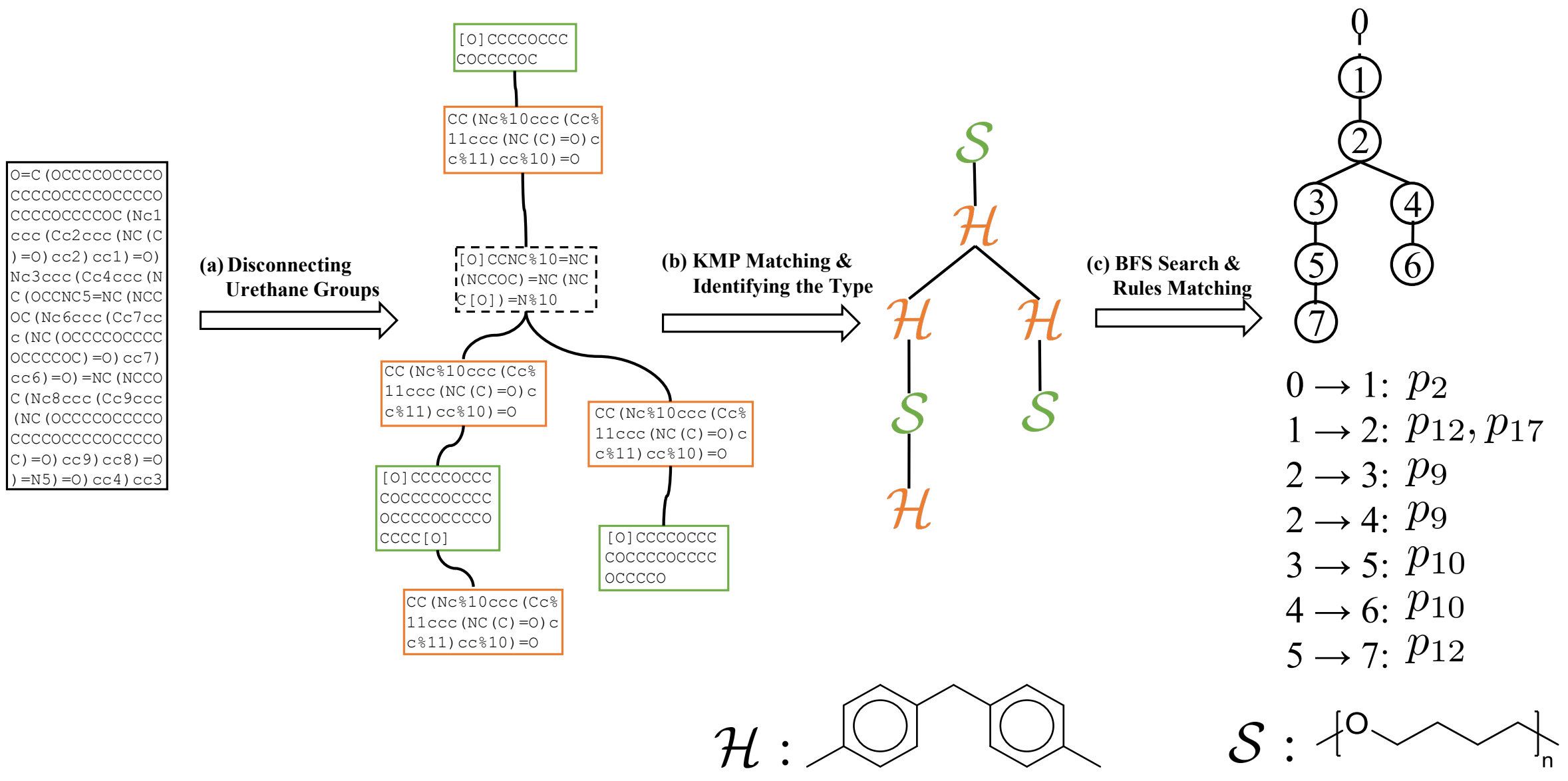

**Figure 10.** Overview of the algorithm for translation from SMILES. The input SMILES string is first broken into a set of molecular fragments, which are identified via string matching. Based on the identity and connections of each fragment, we can construct the hypergraph representation of the molecule. Then, we search for a sequence of PolyGrammar rules that yields the desired result.

This pipeline is sufficient for our needs, but it could be improved with a heuristic search such as $A^*$ search[64], best-first search[65], or learned heuristic search[66] where a heuristic function accelerates the search process by directing attention toward the most promising regions of the search space.

To validate our approach and demonstrate the capacity of our proposed PolyGrammar, we have collected and inversely modelled over 600 polyurethane structures from literature. Many of these polyurethanes are commonly used in synthesis and real-world fabrication, and they feature a wide range of constituent molecules. In particular, the dataset features 8 different types of isocyanates, 11 types of polyols and 7 types of chain extenders. Additional details about our dataset – including information about how to add and translate new polyurethane structures – are described in Supporting Information. Supporting Information also contains several examples polyurethanes from our dataset, which were successfully converted from SMILES to the PolyGrammar representations. Moreover, we emphasize that *each* of the collected SMILES strings in our dataset can be successfully converted to a sequence of production rules in the PolyGrammar. This proves that our PolyGrammar has high representative capacity over a large span of polyurethane structures.

**6. Generalization to Other Polymers**

Our PolyGrammar can also be easily extended to new classes of polymers. These extensions would use the same framework described above, with very few modifications. In the Supporting Information, we illustrate the extended PolyGrammar for different types of copolymers, including alternating copolymers and block copolymers. Note that our PolyGrammar in the main paper can already cover random copolymers, branched copolymers, and graft copolymers. Users only need to add new types of reactants to the symbolic representation in order to determine the species of monomer.

For now, PolyGrammar focus on the backbone structure, i.e., the arrangement of monomers, which largely determines the property of copolymers (derived from more than one species of monomer). The grammar treats the monomer fragment as a whole and distinguishes different monomer types using different symbols. However, there is also a wide range of polymers consisting of only one single type of repeat unit, i.e., homopolymers, where the backbone structures are not variable and the functional group (also called functional residue) of the monomer contributes to the polymer property. To handle this, we augment our PolyGrammar with an additional set of production rules focusing on the representation and generation of functional groups. We also demonstrate the effectiveness of our augmented PolyGrammar using the polyacrylate as an illustrative example. This functional-group grammar together with the basic PolyGrammar (full set of the production rules in Supporting Information) serves as a *hierarchical* generative model for polymers, where the latter one handles the backbone and the former one focuses on the functional residue of each composed monomer. More examples are shown in Supporting Information.

**7. Discussion**

PolyGrammar is an effective chemistry design model that satisfies all five desirable properties discussed in Introduction. In particular, our symbolic representation can convey all possible polyurethane structures in an explicit yet concise manner. The generative model based on this representation is exhaustive (it is capable of generating *any* polyurethane) and trustworthy (every generated polyurethane is guaranteed to be valid). Moreover, the generation process is fully transparent and understandable to the user, as it returns a sequence of meaningful production rules that yield our model's result. Lastly, the generation process is invertible, so molecules can be translated from other popular representations such as SMILES. These superior properties make PolyGrammar more comprehensive and practical than existing representation schemes and generative models. Our full chemical design model (representation, generative model, and inverse model) are also simple and efficient to use in practice. For a polyurethane chain of length 20, the average generation time via PolyGrammar is 4 ms and its translation from SMILES costs 11 ms on a PC with an Intel Core i7 CPU.

For now, our PolyGrammar focuses on single-chained molecular structures. However, real synthesized polyurethanes are a mixture of differently structured chains, where interactions between chains such as hydrogen bonding and crosslinking may occur[47,48]. These interactions influence the physical and chemical property of the polyurethane, largely determining whether the synthesized polyurethane is thermoset or thermoplastic. Such interactions are not currently addressed in PolyGrammar, but they could be added by augmenting the production rules to support interactions between multiple chains.

The current generative model of the PolyGrammar also only imitates the chain-growth polymerization. Although this polymerization mechanism has some benefits for the simulation of polyurethane chains[61], it would be ideal for our PolyGrammar to imitate step-growth polymerization as well. More advanced grammar such as universal grammar[62] will be helpful to achieve this.

These aforementioned features are intriguing and will be implemented and demonstrated in future work. However, even without these augmentations, our proposed PolyGrammar takes an important step toward a more practical and comprehensive system for polymer discovery and exploration.

## 8. Conclusion

In summary, we propose a parametric context-sensitive grammar, called PolyGrammar, for the representation and generation of polymers. The recursive nature of grammar production enables the generation of any polymer chain using only a simple set of production rules. We also implement an algorithm that can transfer a SMILES string of a polymer chain to the sequence of production rules used to generate it. Capable of reproducing a large literature-collected dataset, this algorithm demonstrates the completeness and effectiveness of our PolyGrammar. Our PolyGrammar will benefit the polymer community in several ways. The most immediate contribution is our ability to efficiently generate an exhaustive collection of polymer samples. This corpus could be very powerful in conjunction with other methods (e.g., machine learning) to guide the synthesis of physical polymers and facilitate complex tasks like molecular discovery[2-4] and property optimization[13,14,17]. PolyGrammar is also helpful for the reverse engineering of polymer design and production. Our PolyGrammar serves as a blueprint to construct chemical design models for different classes of chemistries, including both organic and inorganic molecules. Eventually, PolyGrammar could improve chemical communication and exploration, by providing a more efficient and effective representation scheme that is widely suitable for complicated polymers.

**Supporting Information**

Supporting Information is available from the author: production rules of global controllable grammar; collected dataset of polyurethane from literature; examples of translation from SMILES; generalized Polygrammar to other polymers; pseudo-code of translation from SMILES; examples of generated polyurethane chains; examples of branched polyurethane chains; examples of acrylate's functional groups; and abbreviations and acronyms.

## Acknowledgments

**General**: The authors would like to thank Dr. Xingcai Zhang and Dr. Ming Xiao from Harvard University, and Dr. Pengfei Zhang from Qingdao University for their helpful comments.

# Supporting Information

**Polygrammar: Grammar for Digital Polymer Representation and Generation**

*Minghao Guo, Wan Shou, Liane Makatura, Timothy Erps, Michael Foshey, Wojciech Matusik**

## S1. Production Rules of Global Controllable Grammar

The basic idea of the global controllable grammar is to use a message to collect global information about the chain. The message passes back and forth between the left and right ends of the string. It is achieved by swapping the message's position with the adjacent symbol's one at a time. This swapping continues along a certain direction until the message gets to the string end. At each position swap, the message updates its parameters to collect the information required for the control setting. The full set of the production rules for the message passing mechanism is illustrated as follows. Note that in this case, all of the symbols are non-terminal symbols.

| | | | | | |
|---|---|---|---|---|---|
| $p_1$ : | $None$ | $< \mathcal{X} >$ | $None$: | $None$ | $\rightarrow h\mathcal{M}(1,1,0,1)\mathcal{H}h$ |
| $p_2$ : | $None$ | $< \mathcal{X} >$ | $None$: | $None$ | $\rightarrow h\mathcal{H}\mathcal{M}(1,1,0,0)h$ |
| $p_3$ : | $None$ | $< \mathcal{X} >$ | $None$: | $None$ | $\rightarrow s\mathcal{M}(1,0,1,1)\mathcal{S}s$ |
| $p_4$ : | $None$ | $< \mathcal{X} >$ | $None$: | $None$ | $\rightarrow s\mathcal{S}\mathcal{M}(1,0,1,0)s$ |
| $p_5$ : | $None$ | $< lower >$ | $\mathcal{M}(l,r,t,d)$: | $d == 1\ and\ l < L\ and\ r < R$ | $\rightarrow h\mathcal{M}(0,0,0,0)\mathcal{H}$ |
| $p_6$ : | $None$ | $< lower >$ | $\mathcal{M}(l,r,t,d)$: | $d == 1\ and\ l < L$ | $\rightarrow s\mathcal{M}(0,0,1,0)\mathcal{S}$ |
| $p_7$ : | $lower$ | $< \mathcal{M}(l,r,t,d) >$ | $upper$: | $d == 1$ | $\rightarrow Null$ |
| $p_8$ : | $\mathcal{M}(l,r,t,d)$ | $< \mathcal{H} >$ | $\mathcal{H}$: | $d == 0$ | $\rightarrow \mathcal{M}\left(l+1, \frac{r*l+1}{l+1}, 0, d\right)$ |
| $p_9$ : | $\mathcal{M}(l,r,t,d)$ | $< \mathcal{H} >$ | $h$: | $d == 0$ | $\rightarrow \mathcal{M}\left(l+1, \frac{r*l+1}{l+1}, 0, d\right)$ |
| $p_{10}$ : | $\mathcal{M}(l,r,t,d)$ | $< \mathcal{H} >$ | $\mathcal{S}$: | $d == 0$ | $\rightarrow \mathcal{M}\left(l+1, \frac{r*l+1}{l+1}, 1, d\right)$ |
| $p_{11}$ : | $\mathcal{M}(l,r,t,d)$ | $< \mathcal{S} >$ | $\mathcal{H}$: | $d == 0$ | $\rightarrow \mathcal{M}\left(l+1, \frac{r*l}{l+1}, 0, d\right)$ |
| $p_{12}$ : | $\mathcal{M}(l,r,t,d)$ | $< \mathcal{S} >$ | $\mathcal{S}$: | $d == 0$ | $\rightarrow \mathcal{M}\left(l+1, \frac{r*l}{l+1}, 1, d\right)$ |
| $p_{13}$ : | $\mathcal{M}(l,r,t,d)$ | $< \mathcal{S} >$ | $s$: | $d == 0$ | $\rightarrow \mathcal{M}\left(l+1, \frac{r*l}{l+1}, 1, d\right)$ |
| $p_{14}$ : | $lower$ | $< \mathcal{M}(l,r,t,d) >$ | $upper$: | $d == 0\ and\ t == 1$ | $\rightarrow \mathcal{S}$ |
| $p_{15}$ : | $lower$ | $< \mathcal{M}(l,r,t,d) >$ | $upper$: | $d == 0\ and\ t == 1$ | $\rightarrow \mathcal{H}$ |
| $p_{16}$ : | $upper$ | $< \mathcal{M}(l,r,t,d) >$ | $upper$: | $t == 1$ | $\rightarrow \mathcal{S}$ |
| $p_{17}$ : | $upper$ | $< \mathcal{M}(l,r,t,d) >$ | $upper$: | $t == 0$ | $\rightarrow \mathcal{H}$ |

$p_{18}: \mathcal{M}(l,r,t,d) \quad < lower > \quad None: \quad d == 1 \; and \; l < L \; and \; r < R \rightarrow \mathcal{HM}(0,0,0,1)h$

$p_{19}: \mathcal{M}(l,r,t,d) \quad < lower > \quad None: \quad d == 1 \; and \; l < L \quad \rightarrow \mathcal{SM}(0,0,1,1)s$

$p_{20}: upper \; < \mathcal{M}(l,r,t,d) > \quad lower: \quad d == 0 \quad \rightarrow Null$

$p_{21}: \mathcal{H} \quad < \mathcal{H} > \quad \mathcal{M}(l,r,t,d): \quad d == 1 \quad \rightarrow \mathcal{M}\left(l+1, \frac{r*l+1}{l+1}, 0, d\right)$

$p_{22}: h \quad < \mathcal{H} > \quad \mathcal{M}(l,r,t,d): \quad d == 1 \quad \rightarrow \mathcal{M}\left(l+1, \frac{r*l+1}{l+1}, 0, d\right)$

$p_{23}: \mathcal{S} \quad < \mathcal{H} > \quad \mathcal{M}(l,r,t,d): \quad d == 1 \quad \rightarrow \mathcal{M}\left(l+1, \frac{r*l+1}{l+1}, 1, d\right)$

$p_{24}: \mathcal{H} \quad < \mathcal{H} > \quad \mathcal{M}(l,r,t,d): \quad d == 1 \quad \rightarrow \mathcal{M}\left(l+1, \frac{r*l}{l+1}, 0, d\right)$

$p_{25}: \mathcal{S} \quad < \mathcal{H} > \quad \mathcal{M}(l,r,t,d): \quad d == 1 \quad \rightarrow \mathcal{M}\left(l+1, \frac{r*l}{l+1}, 1, d\right)$

$p_{26}: s \quad < \mathcal{H} > \quad \mathcal{M}(l,r,t,d): \quad d == 1 \quad \rightarrow \mathcal{M}\left(l+1, \frac{r*l}{l+1}, 1, d\right)$

$p_{27}: upper \; < \mathcal{M}(l,r,t,d) > \quad lower: \quad d == 1 \; and \; t == 1 \quad \rightarrow \mathcal{S}$

$p_{28}: upper \; < \mathcal{M}(l,r,t,d) > \quad lower: \quad d == 1 \; and \; t == 0 \quad \rightarrow \mathcal{H}$

Similar to the PolyGrammar in the main paper, the global controllable grammar is also a context-sensitive parametric grammar. $\mathcal{M}$ denotes the message, $lower$ indicates the lower case symbols, containing $h$ and $s$, $upper$ indicates the upper case symbols, containing $\mathcal{H}$ and $\mathcal{S}$. For each production rule in the grammar, the "→" separates the predecessor and the successor. The symbol to be replaced is inside the "< >". The contexts are the symbols located at both sides of "< >" in the predecessor ($None$ indicates no constraints). Parameters of message symbol $\mathcal{M}$ locates inside "( )". The logic condition of the parameters for each production rule is between ":" and "→" ($None$ also indicates no constraints here). During the production process, the production rule can only be applied to a symbol when both of its context and the logic condition are satisfied. The production process will stop when no production rules can be invoked, i.e. for each symbol of the string, there are no production rules that can meet the condition and the contexts of the symbol.

With this set of production rules, we can control two constraints of the polyurethane chain: the chain length $L$ and the ratio $R$ of hard segment to soft segment. The message symbol $\mathcal{M}$ propagates back and forth between the left and right end of the string, collects global information of the chain and determines how to expand the string meeting the constraints. The propagation is achieved by switching the message's position with the adjacent symbol's along a certain direction one at a time until the message gets to the string end. At each swap, the message updates its parameters to collect information needed for the control setting. There are four parameters in $\mathcal{M}$ in total: $l$ indicates the current chain length, $r$ indicates the current chain's ratio of hard segment to soft segment, $t$ is an auxiliary parameter to record the symbol that $\mathcal{M}$ is switching with (0 for $\mathcal{H}$, 1 for $\mathcal{S}$), and $d$ indicates the direction that the message is propagating (0 for right, 1 for left). As for the roles that each production rule serves, $p_1$, $p_2$, $p_3$ and $p_4$ initialize the start symbol $\mathcal{X}$. Taking $p_5$, $p_6$ and $p_7$ together, when the left-propagating message arrives at the left end, the string is expanded according to the collected information and the constraints. Meanwhile, this left-propagating message disappears, and a right-propagating message is generated. This right-propagating message continually switches its position with its right-neighbor symbol using the rules $p_8$, $p_9$, $p_{10}$, $p_{11}$, $p_{12}$, $p_{13}$, $p_{14}$, $p_{15}$, $p_{16}$ and $p_{17}$. Similarly, $p_{18}$, $p_{19}$ and $p_{20}$ expand the string when the right-propagating message arrives at the right end and generate a left-propagating message. $p_{21}$, $p_{22}$, $p_{23}$, $p_{24}$, $p_{25}$, $p_{26}$, $p_{27}$, $p_{28}$, $p_{16}$ and $p_{17}$ propagate the left-propagating message by iteratively switching its position with its left-neighbor symbol. A detailed illustrative example of the message passing mechanism is shown in Figure. S1.

The example in Figure. S1 illustrates the production process under the constraints $L = 3$, $R = 0.67$, where the chain length is 3 and the ratio of hard segment to soft segment is lower than 0.7. At step 1, $p_1$ expands the initial symbol $\mathcal{X}$. The parameters of $\mathcal{M}$ indicate that the message is currently propagating along the left direction ($d = 1$). Since it has already reached the left end of the string and the current ratio $r = 1$ is larger than the constraint $R = 0.67$, at step 2, $p_6$ expands the left end of the string with $\mathcal{S}$ and also generates a new message $\mathcal{M}$ ($l = 0, r = 0$) propagating along the right direction ($d = 0$). Meanwhile, the previous message disappears by $p_7$. Then at step 3 and step 4, the message propagates along the right direction by switching position with its right neighbor. The switching is assisted with the auxiliary parameter $t$. Also during the propagation, $l$ and $r$ are updated to record the information of chain length and the ratio of hard segment to soft segment. At step 5, the message reaches the right end of the string, so $p_{18}$ expands the string with $\mathcal{H}$ as the ratio $r = 0.5$ is smaller than $R = 0.67$. A new left-propagating message is generated and the old message disappears. At step 6, 7, and 8, the message carries on the propagation and collects the information. At step 9, when the message reaches the left end, since the chain length $l = 3$ meets the constraints $L = 3$, the message disappears and the production process concludes.

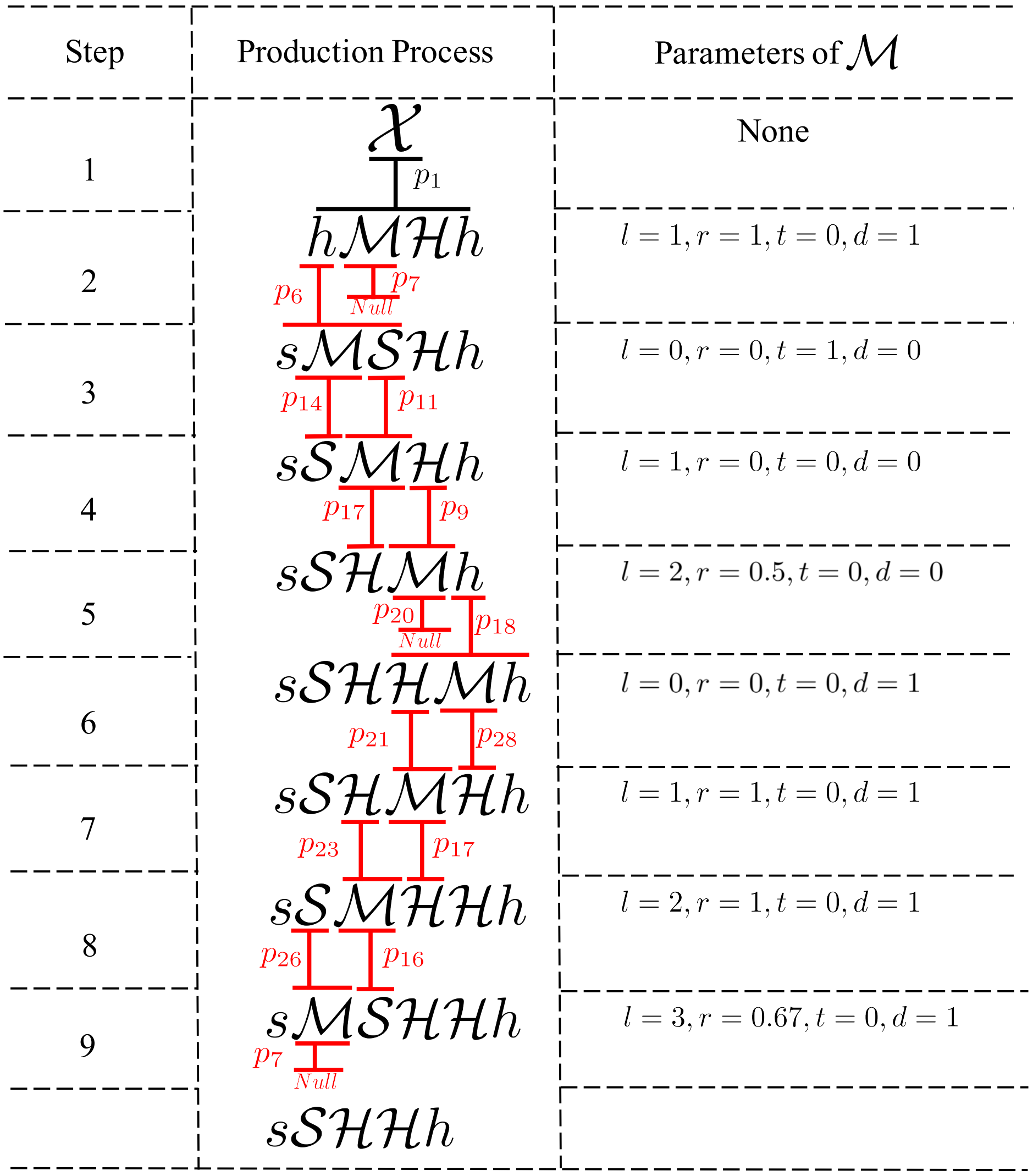

**Figure. S1.** Illustrative example of the message passing mechanism with constraints $L = 3$, $R = 0.67$.

## S2. Collected Dataset of Polyurethane from Literature

We have collected a dataset of polyurethane data from literature, including 8 different types of isocyanates, 11 types of polyols and 7 types of chain extenders. Each sample is illustrated in the form of BigSMILES. Detailed samples are illustrated as follows,

| Name | BigSMILES |
|---|---|
| Diisocynates | |
| TDI | CC1=CC=C(NC(=O)>)C=C1NC(=O)> |
| MDI | >C(=O)Nc1ccc(Cc2cccNC(=O)>cc2)cc1 |
| HDI | >C(=O)NCCCCCCNC(=O)> |
| IPDI | CC1(C)CC(NC(=O)>)CC(CNC(=O)>)(C)C1 |
| DBDI | >C(=O)Nc1ccc(CCc2ccc(NC(=O)>)cc2)cc1 |
| HMDI | >C(=O)NC1CCC(CC2CCC(NC(=O)>)CC2)CC1 |
| NDI | >C(=O)Nc1cccc2c(NC(=O)>)cccc12 |
| TMDI | CC(CCNC(=O)>)CC(C)(C)CNC(=O)> |
| Polyols | |
| PTMO/ PTHF | <OCCC{[<]OCCCC[>]}O< |
| PEG/ PEO | <OCC{[<]OCC[>]}O< |
| PEA | <OCCOC(=O)CCCCC(=O){[<]OCCOC(=O)CCCCC(=O)[>]}O< |
| PBA | <OCCCCOC(=O)CCCCC(=O){[<]OCCCCOC(=O)CCCCC(=O)[>]}O< |
| PBU | CC=CC |
| PCL/ PCD | <OCCCCCC(=O)OCCCCCC(=O){[<]OCCCCCC(=O)OCCCCCC(=O)[>]}O< |
| PHA | <OCCCCCCCOC(=O)OCCCCCCCOC(=O)O{[<]OCCCCCCCOC(=O)OCCCCCCCOC(=O)O[>]}O< |
| PET | <OC(=O)c1ccc(cc1)C(=O)OCC{[<]OC(=O)c1ccc(cc1)C(=O)OCC[>]}O< |
| PLA | <OC(C)C(=O){[<]OC(C)C(=O)[>]}O< |
| CHDM | <OCC1CCC(CC1)C{[<]OCC1CCC(CC1)C[>]}O< |
| | |
| Poly bd | <OCC=CCCC(C=C)CC=CC{[<]OCC=CCCC(C=C)CC=CC}O< |
| Chain Extenders | |
| BDO/<br>BD/BG | <OCCCCO< |
| EG | <OCCO< |
| DEG | <OCCOCCO< |
| DAPO | <Nc1ccc(cc1)-c2nnc(o2)-c3ccc(cc3)N< |
| DAB | <Nc1ccc(CCc2ccc(N<)cc2)cc1 |
| DAPy | C1=CC(=NC(=C1)N(<))N< |
| MDA | <Nc1ccc(Cc2ccc(N<)cc2)cc1 |

By combining 3 types of components, this dataset contains $8 \times 11 \times 7 = 616$ types of polyurethanes that are commonly used in the synthesis and real-world fabrication. The full names of the abbreviations in the dataset are listed in Table S5. These data samples are stored in a “.CSV” file and can be easily handled using Python code to perform the algorithms of generative model and translation from SMILES. It is also capable to add new structures to this dataset. The only thing to do is to convert the structure to the BigSMILES format and add it to the “.CSV” file.

## S3. Examples of Translation from SMILES

1. **Input SMILES:**

   CC1(C)CC(NC(=O)OCCCCOCCCCOCCCCOCCCCOC(=O)NCC2(C)CC(NC(=O)OCCCCOC(=O)NCC3(C)CC(NC(=O)OCCCCOC(=O)NCC4(C)CC(NC(=O)OCCCCOCCCCOC(=O)NCC5(C)CC(NC(=O)OCCCCOC(=O)NCC6(C)CC(NC(=O)OCCCCOCCCCOCCCCOCCCCOCCCCO)CC(C)(C)C6)CC(C)(C)C5)CC(C)(C)C4)CC(C)(C)C3)CC(C)(C)C2)CC(C)(CN=C=O)C1

   **Translation Results:**

   Component types: IPDI, PTMO, BDO

Symbolic hypergraph string: $\mathcal{HSHHHSHHS}$

Production rules: $\{p_1, p_{10}, p_{12}, p_9, p_9, p_{10}, p_{12}, p_9, p_{10}, p_5, p_{14}\}$

2. **Input SMILES:**

```
Cc1ccc(NC(=O)OCCCCOC(=O)Nc2cc(NC(=O)OCCCCOCCCCOCCCCOCCCCO
CCCCOCCCCOCCCCOC(=O)Nc3cc(NC(=O)OCCOC(=O)Nc4cc(NC(=O)OCCO
C(=O)Nc5cc(NC(=O)OCCCCOC(=O)Nc6cc(NC(=O)OCCCCO)ccc6C)ccc5
C)ccc4C)ccc3C)ccc2C)cc1NC(=O)OCCCCO
```

**Translation Results:**

Component types: TDI, PTMO, EG

Symbolic hypergraph string: $\mathcal{SHSHHHSHSHS}$

Production rules: $\{p_1, p_4, p_{10}, p_6, p_4, p_6, p_3, p_3, p_4, p_6, p_4, p_8, p_{14}\}$

3. **Input SMILES:**

```
CC(OC(=O)Nc1cccc2c(NC(=O)OCCCCOC(=O)Nc3cccc4c(NC(=O)OCCCC
OC(=O)Nc5cccc6c(NC(=O)OC(=O)C(C)OC(=O)C(C)OC(=O)Nc7cccc8c
(NC(=O)OC(=O)C(C)OC(=O)Nc9cccc%10c(NC(=O)OCCCCOC(=O)Nc%11
cccc%12c(N=C=O)cccc%11%12)cccc9%10)cccc78)cccc56)cccc34)c
ccc12)C(=O)OC(=O)Nc1cccc2c(NC(=O)OCCCCOC(=O)Nc3cccc4c(NC(
=O)OC(C)C(=O)OC(=O)Nc5cccc6c(NC(=O)OCCCCOC(=O)Nc7cccc8c(N
C(=O)OCCCCOC(=O)Nc9cccc%10c(NC(=O)OCCCCOC(=O)Nc%11cccc%12
c(N=C=O)cccc%11%12)cccc9%10)cccc78)cccc56)cccc34)cccc12
```

**Translation Results:**

Component types: NDI, PLA, BDO

Symbolic hypergraph string: $\mathcal{HHSHSHHHSHHSHHHH}$

Production rules: $\{p_2, p_6, p_{12}, p_3, p_9, p_3, p_{10}, p_4, p_{12}, p_6, p_9, p_4, p_9, p_6, p_9, p_3, p_5, p_{11}\}$

4. **Input SMILES:**

O=C(NC1CCC(CC2CCC(NC(=O)OCCOCCOC(=O)NC3CCC(CC4CCC(NC(=O)OCCOC(=O)NC5CCC(CC6CCC(NC(=O)OCCOC(=O)NC7CCC(CC8CCC(NC(=O)OCCOCCOC(=O)NC9CCC(CC%10CCC(NC(=O)OCCOCCOC(=O)NC%11CCC(CC%12CCC(NC(=O)OCCOCCOC(=O)NC%13CCC(CC%14CCC(NC(=O)OCCOC(=O)NC%15CCC(CC%16CCC(NC(=O)OCCOCCO)CC%16)CC%15)CC%14)CC%13)CC%12)CC%11)CC%10)CC9)CC8)CC7)CC6)CC5)CC4)CC3)CC2)CC1)OCCOCCOCCOCCO

**Translation Results:**

Component types: HMDI, PEG, DEG

Symbolic hypergraph string: $\mathcal{SHHHHHHHHHHHS}$

Production rules: $\{p_1, p_{10}, p_3, p_3, p_3, p_3, p_3, p_3, p_3, p_3, p_3, p_3, p_4, p_8, p_{14}\}$

5. **Input SMILES:**

O=C=Nc1ccc(Cc2ccc(NC(=O)OCCCCOCCCCOCCCCOC(=O)Nc3ccc(Cc4ccc(NC(=O)OCCOCCOC(=O)Nc5ccc(Cc6ccc(NC(=O)OCCCCOC(=O)Nc7ccc(Cc8ccc(NC(=O)OCCOCCOC(=O)Nc9ccc(Cc%10ccc(NC(=O)OCCCCOCCCCOCCCCOCCCCOCCCCOC(=O)Nc%11ccc(Cc%12ccc(NC(=O)OCCCCOC(=O)Nc%13ccc(Cc%14ccc(N=C=O)cc%14)cc%13)cc%12)cc%11)cc%10)cc9)cc8)cc7)cc6)cc5)cc4)cc3)cc2)cc1

**Translation Results:**

Component types: MDI, PTMO, DEG

Symbolic hypergraph string: $\mathcal{HSHHSHHSHSH}$

Production rules: $\{p_1, p_{10}, p_{12}, p_9, p_{10}, p_{12}, p_9, p_{10}, p_{12}, p_{10}, p_{12}, p_5, p_{11}\}$

6. **Input SMILES:**

O=C=NCCCCCCNC(=O)OCCCCCC(=O)OCCCCCC(=O)OCCCCCC(=O)OCCCCCC(=O)OCCCCCC(=O)OCCCCCC(=O)OCCCCCC(=O)OCCCCCC(=O)OCCCCCC(=

O)OCCCCCC(=O)OCCCCCC(=O)OCCCCCC(=O)OC(=O)NCCCCCCNC(=O)OCC
OC(=O)NCCCCCCNC(=O)OCCOC(=O)NCCCCCCNC(=O)OCCCCCC(=O)OCCCC
CC(=O)OCCCCCC(=O)OCCCCCC(=O)OCCCCCC(=O)OCCCCCC(=O)OC(=O)N
CCCCCCNC(=O)OCCCCCC(=O)OCCCCCC(=O)OC(=O)NCCCCCCNC(=O)OCCO
C(=O)NCCCCCCN=C=O

**Translation Results:**

Component types: HDI, PCL, EG

Symbolic hypergraph string: $\mathcal{HSHHHSHSHH}$

Production rules: $\{p_1, p_{10}, p_{12}, p_9, p_9, p_{10}, p_{12}, p_{10}, p_{12}, p_9, p_5, p_{11}\}$

7. **Input SMILES:**

CC(O)C(=O)OC(C)C(=O)OC(=O)Nc1ccc(Cc2ccc(NC(=O)OC(C)C(=O)O
C(=O)Nc3ccc(Cc4ccc(NC(=O)OCCCCOC(=O)Nc5ccc(Cc6ccc(NC(=O)O
CCCCOC(=O)Nc7ccc(Cc8ccc(NC(=O)OC(C)C(=O)OC(C)C(=O)OC(C)C(
=O)OC(=O)Nc9ccc(Cc%10ccc(NC(=O)OC(C)C(=O)OC(=O)Nc%11ccc(C
c%12ccc(NC(=O)OC(C)C(=O)OC(=O)Nc%13ccc(Cc%14ccc(NC(=O)OCC
CCOC(=O)Nc%15ccc(Cc%16ccc(NC(=O)OC(C)C(=O)O)cc%16)cc%15)c
c%14)cc%13)cc%12)cc%11)cc%10)cc9)cc8)cc7)cc6)cc5)cc4)cc3)
cc2)cc1

**Translation Results:**

Component types: MDI, PLA, BDO

Symbolic hypergraph string: $\mathcal{SHSHHHSHSHSHHS}$

Production rules: $\{p_2, p_{12}, p_{10}, p_{12}, p_9, p_9, p_{10}, p_{12}, p_{10}, p_{12}, p_{10}, p_{12}, p_9, p_{10}, p_8, p_{14}\}$

8. **Input SMILES:**

C=CC(CC=CO)CCC=CCOC=CCC(C=C)CCC=CCOC=CCC(C=C)CCC=CCOC=CCC
(C=C)CCC=CCOC(=O)NC1CC(C)(C)CC(C)(CNC(=O)OC=CCC(C=C)CCC=C

COC(=O)NC2CC(C)(C)CC(C)(CNC(=O)Nc3cccc(NC(=O)NC4CC(C)(C)C
C(C)(CNC(=O)OC=CCC(C=C)CCC=CCOC=CCC(C=C)CCC=CCOC=CCC(C=C)
CCC=CCOC(=O)NC5CC(C)(C)CC(C)(CNC(=O)OC=CCC(C=C)CCC=CCOC=C
CC(C=C)CCC=CCOC(=O)NC6CC(C)(C)CC(C)(CNC(=O)NC7CCCC(NC(=O)
NC8CC(C)(C)CC(C)(CNC(=O)Nc9cccc(NC(=O)NC%10CC(C)(C)CC(C)(
CN=C=O)C%10)n9)C8)n7)C6)C5)C4)n3)C2)C1

**Translation Results:**

Component types: IPDI, Poly bd, DAPy

Symbolic hypergraph string: $\mathcal{SHSHS}$

Production rules: $\{p_2, p_{12}, p_{10}, p_{13}, p_{12}, p_{10}, p_{13}, p_{13}, p_8, p_{14}\}$

**9. Input SMILES:**

C=CC(CC=COCC=CCCC(C=C)CC=COC(=O)Nc1ccc(Cc2ccc(NC(=O)OCCOC
COC(=O)Nc3ccc(Cc4ccc(NC(=O)OCC=CCCC(C=C)CC=COC(=O)Nc5ccc(
Cc6ccc(NC(=O)OCC=CCCC(C=C)CC=COC(=O)Nc7ccc(Cc8ccc(NC(=O)O
CC=CCCC(C=C)CC=COC(=O)Nc9ccc(Cc%10ccc(NC(=O)OCCOCCOC(=O)N
c%11ccc(Cc%12ccc(NC(=O)OCC=CCCC(C=C)CC=COC(=O)Nc%13ccc(Cc
%14ccc(NC(=O)OCC=CCCC(C=C)CC=COC(=O)Nc%15ccc(Cc%16ccc(N=C
=O)cc%16)cc%15)cc%14)cc%13)cc%12)cc%11)cc%10)cc9)cc8)cc7)
cc6)cc5)cc4)cc3)cc2)cc1)CCC=CCOC=CCC(C=C)CCC=CCOC(=O)Nc1c
cc(Cc2ccc(N=C=O)cc2)cc1

**Translation Results:**

Component types: MDI, Poly bd, DEG

Symbolic hypergraph string: $\mathcal{HSHSHHSHSHSHHSH}$

Production rules: $\{p_2, p_6, p_{12}, p_3, p_4, p_6, p_4, p_6, p_4, p_6, p_3, p_4, p_6, p_4, p_6, p_5, p_{11}\}$

**10. Input SMILES:**

O=C(O)CCCCC(=O)OCCOC(=O)Nc1cccc2c(NC(=O)OC(=O)CCCCC(=O)OCCOC(=O)Nc3cccc4c(NC(=O)OC(=O)CCCCC(=O)OCCOC(=O)CCCCC(=O)OCCOC(=O)Nc5cccc6c(NC(=O)Nc7cccc(NC(=O)Nc8cccc9c(NC(=O)Nc%10cccc(NC(=O)Nc%11cccc%12c(NC(=O)Nc%13cccc(NC(=O)Nc%14cccc%15c(NC(=O)Nc%16cccc(NC(=O)Nc%17cccc%18c(NC(=O)Nc%19cccc(NC(=O)Nc%20cccc%21c(NC(=O)OC(=O)CCCCC(=O)OCCOC(=O)Nc%22cccc%23c(NC(=O)OC(=O)CCCCC(=O)OCCOC(=O)CCCCC(=O)OCCOC(=O)CCCCC(=O)OCCO)cccc%22%23)cccc%20%21)n%19)cccc%17%18)n%16)cccc%14%15)n%13)cccc%11%12)n%10)cccc89)n7)cccc56)cccc34)cccc12

**Translation Results:**

Component types: NDI, PEA, DAPy

Symbolic hypergraph string: $\mathcal{S}\mathcal{H}\mathcal{S}\mathcal{H}\mathcal{S}\mathcal{H}\mathcal{S}$

Production rules: $\{p_2, p_{12}, p_{10}, p_{12}, p_{10}, p_{13}, p_{13}, p_{13}, p_{13}, p_{13}, p_{12}, p_{10}, p_8, p_{14}\}$

## S4. Generalized PolyGrammar to Other Polymers

The extended PolyGrammar for different types of copolymers and functional groups is illustrated as follows,

- for block copolymers,

$$p_1 : None < \mathcal{X} > None : None \rightarrow h\mathcal{H}(0.5L_H)h$$
$$p_2 : None < \mathcal{X} > None : None \rightarrow s\mathcal{S}(0.5L_S)s$$
$$p_3 : None < h > \mathcal{H}(x) : x \geq 1 \rightarrow h\mathcal{H}(x-1)$$
$$p_4 : None < h > \mathcal{H}(x) : x < 1 \rightarrow s\mathcal{S}(L_S)$$
$$p_5 : None < h > \mathcal{H}(x) : x < 1 \rightarrow Null$$
$$p_6 : None < s > \mathcal{S}(x) : x \geq 1 \rightarrow s\mathcal{S}(x-1)$$
$$p_7 : None < s > \mathcal{S}(x) : x < 1 \rightarrow h\mathcal{H}(L_H)$$
$$p_8 : None < s > \mathcal{S}(x) : x < 1 \rightarrow Null$$
$$p_9 : \mathcal{H}(x) < h > None : x \geq 1 \rightarrow \mathcal{H}(x-1)h$$

$$p_{10}\text{: } \mathcal{H}(x) < h > None\text{: } x < 1 \rightarrow \mathcal{S}(L_S)s$$
$$p_{11}\text{: } \mathcal{H}(x) < h > None\text{: } x < 1 \rightarrow Null$$
$$p_{12}\text{: } \mathcal{S}(x) < s > None\text{: } x \geq 1 \rightarrow \mathcal{S}(x-1)s$$
$$p_{13}\text{: } \mathcal{S}(x) < s > None\text{: } x < 1 \rightarrow \mathcal{H}(L_H)h$$
$$p_{14}\text{: } \mathcal{S}(x) < s > None\text{: } x < 1 \rightarrow Null$$

- for alternating copolymers,

$$p_1\text{ : } None < \mathcal{X} > None\text{: } None \rightarrow h\mathcal{H}(L)h$$
$$p_2\text{ : } None < \mathcal{X} > None\text{: } None \rightarrow s\mathcal{S}(L)s$$
$$p_3\text{ : } None < h > \mathcal{H}(x)\text{: } x \geq 1 \rightarrow s\mathcal{S}(x-1)$$
$$p_4\text{ : } None < h > \mathcal{H}(x)\text{: } x < 1 \rightarrow Null$$
$$p_5\text{ : } None < s > \mathcal{S}(x)\text{: } x \geq 1 \rightarrow h\mathcal{H}(x-1)$$
$$p_6\text{ : } None < s > \mathcal{S}(x)\text{: } x < 1 \rightarrow Null$$
$$p_7\text{ : } \mathcal{H}(x) < h > None\text{: } x \geq 1 \rightarrow \mathcal{S}(x-1)s$$
$$p_8\text{ : } \mathcal{H}(x) < h > None\text{: } x < 1 \rightarrow Null$$
$$p_9\text{ : } \mathcal{S}(x) < s > None\text{: } x \geq 1 \rightarrow \mathcal{H}(x-1)h$$
$$p_{10}\text{: } \mathcal{S}(x) < s > None\text{: } x < 1 \rightarrow Null$$

- for functional groups of polyacrylates,

$$p_1\text{ : } b(x) < \mathcal{F} > None\text{: } y = x + z \rightarrow c(y)\mathcal{A}(z)$$
$$p_2\text{ : } None < \mathcal{A}(z) > None\text{: } z > 1 \rightarrow [\mathcal{A}(z-1)]\mathcal{A}(1)$$
$$p_3\text{ : } None < \mathcal{A}(z) > None\text{: } z \geq 1 \rightarrow [b(x)]\mathcal{F}$$
$$p_4\text{ : } None < \mathcal{A}(z) > None\text{: } z \leq 1 \rightarrow None$$

## S5. Pseudo-code of Translation from SMILES

The inverse design process contains three parts: disconnecting carbamate bonds, constructing the hypergraph and BFS searching for rules matching. The pseudo code is illustrated as follows.

---

**Algorithm 1:** Pseudo-code of translation from SMILES.

**Input:**

SMILES string of polyurethane chain $\mathcal{P}$

The set of production rules of PolyGrammar $\{p_i \mid i = 1, \dots, N\}$

The set of SMILES strings of isocyanate candidates $\{\mathcal{I}_i \mid i = 1, \dots, N_J\}$

The set of SMILES strings of macrodiol candidates $\{\mathcal{D}_i | \ i = 1, \ldots, N_D\}$
The set of SMILES strings of chain-extender candidates $\{\mathcal{C}_i | \ i = 1, \ldots, N_C\}$

**Output:**

The sequence of the production rules $\{p_k | k = i_i, \ldots, i_K\}$ to produce $\mathcal{P}$

**Function** GetSubstructMatches($a$, $s$):

```
/*
input:
    SMILES string, a (the substructure sought)
    SMILES string, s (the total molecule to be searched)
output:
    an array of integers or ∅, P (positions in s at which a is found)
*/
/* standard substructure matching algorithm */
```

**return** $P$;

**Function** KMPMatching($a$, $s$):

```
/*
input:
    SMILES string, a (the word sought)
    SMILES string, s (the text to be searched)
output:
    an array of integers, P positions in s at which a is found)
*/
/* standard KMP string matching algorithm */
```

**return** $P$;

**Function** ConstructGraph($a$, $n$):

```
/*
input:
    adjacency matrix, a
    an array of nodes, n
output:
    an undirected graph, G
*/
/* standard undirected graph construction algorithm */
```

**return** $G$;

**Function** BFSearch($G$):

```
/*
input:
    an undirected graph, G
output:
    an array of traversed edges, e
*/
/* standard Breadth-first Search algorithm */
```

**return** $e$;

```
/* 1. Disconnecting carbamate bonds */
```

$CBondSMILES \leftarrow$ SMILES of $Carbamate\ Bond$;
$CBondIdx \leftarrow$ GetSubstructMatches($CBondSMILES, \mathcal{P}$);
$n \leftarrow$ length($CBondIdx$);

```
FragmentSet ← ∅;
CBondSet ← ∅;
StartIdx ← 0;
for i ← 0 to n do
    if i == n then
        location ← i;
    else
        EndIdx ← CBondIdx[i];
        FragmentSet ← FragmentSet ∪ {P[StartIdx: EndIdx]};
        BondEndIdx ← EndIdx + length(CBond);
        CBondSet ← CBondSet ∪ {P[EndIdx: BondEndIdx]};
        StartIdx ← BondEndIdx;
/* 2. Identifying the symbol type and constructing the hypergraph */
n ← length(FragmentSet);
AdjacencyMatrix ← n × n zero matrix ;
HyperNodes ← ∅;
/* get the adjacency matrix */
for i ← 0 to n − 1 do
    CBond ← CBondSet[i];
    CBondStart ← CBond[0];
    CBondEnd ← CBond[end];
    for j ← 0 to n − 1 do
        Fragment ← FragmentSet[i];
        if CBondStart in Fragment then
            StartIdx = j;
        else if CBondEnd in Fragment then
            EndIdx = j;
    AdjacencyMatrix[StartIdx][EndIdx] = 1;
    AdjacencyMatrix[EndIdx][StartIdx] = 1;
/* get the symbol type */
for i ← 0 to n − 1 do
    if {KMPMatching(Fragment, I_i)} is not ∅ then
        HyperNodes ← HyperNodes ∪ {'H'};
    else if {KMPMatching(Fragment, D_i)} is not ∅ then
        HyperNodes ← HyperNodes ∪ {'S'};
    else if {KMPMatching(Fragment, C_i)} is not ∅ then
        HyperNodes ← HyperNodes ∪ {'u'};
HyperGraph ← ConstructGraph(AdjacencyMatrix, HyperNodes);
/* 3. BFS searching the graph and rules matching */
EdgeList ← BFSearch(HyperGraph);
P ← ∅;
n ← length(EdgeList);
for i ← 0 to n − 1 do
    Edge ← EdgeList[i];
    Pre ← Edge.predecessor;
    Suc ← Edge.successor;
    Rule ← {p_i}.find(<Pre, Suc>);
    P ← P ∪ Rule;
return P
```

## S6. Examples of Generated Polyurethane Chains

| Index | Diisocyanate | Macrodiol | Chain Extender | Generated Hypergraph Symbolic String |
|---|---|---|---|---|
| 1 | DBDI | PTMO | DAPO | $\mathcal{HSHHSHHHHSHHHHHH}$ |
| | 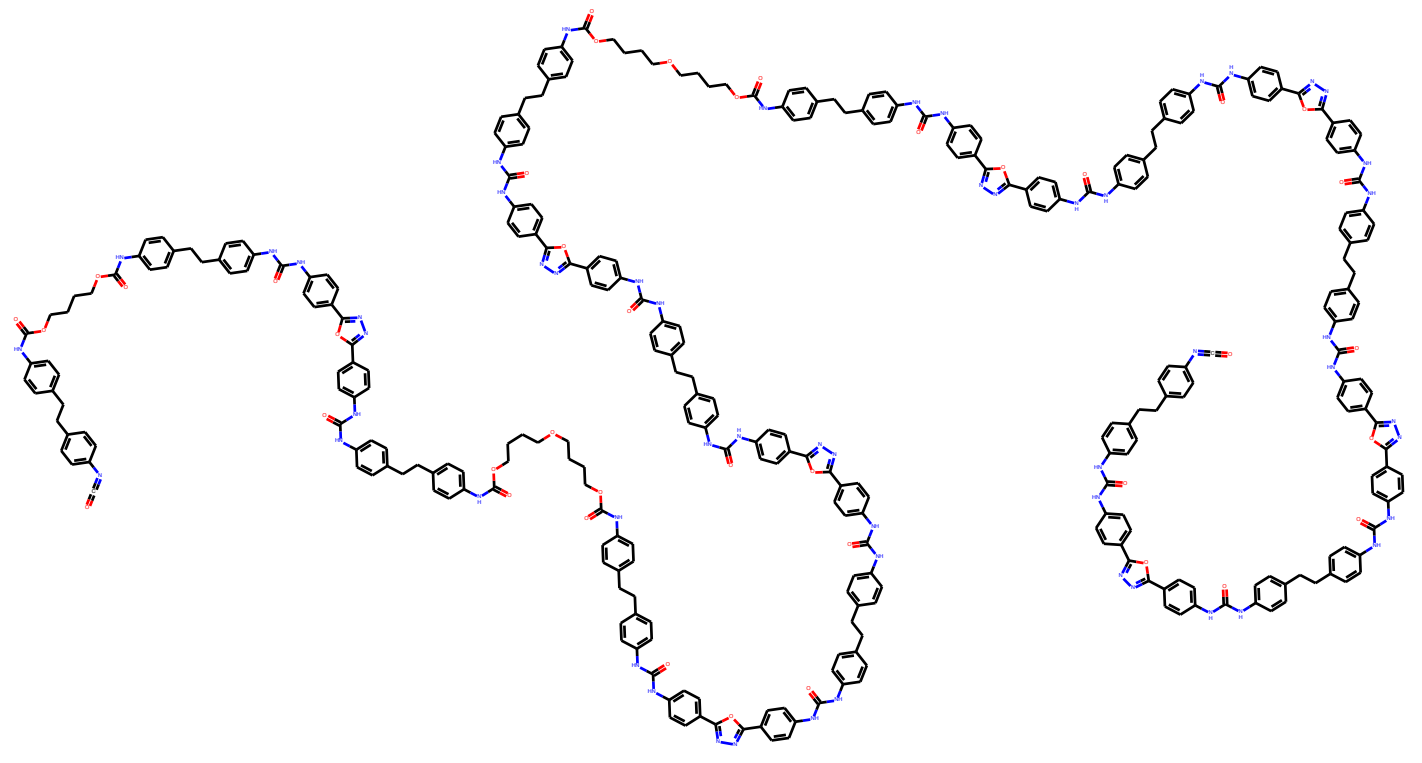 | | | |
| 2 | TMDI | PCL | DAB | $\mathcal{HSHHSHHSHSHHHS}$ |
| | 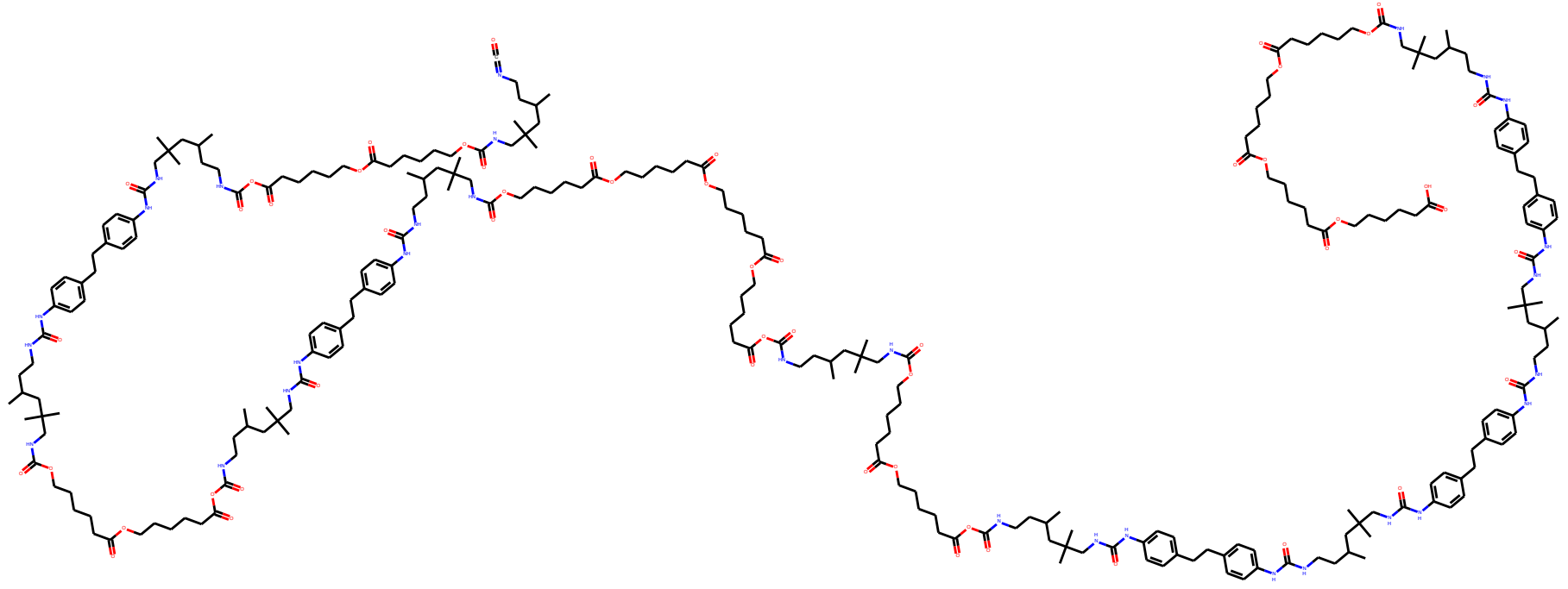 | | | |
| 3 | NDI | PEA | DAPy | $\mathcal{SHSHHSHHS}$ |
| | 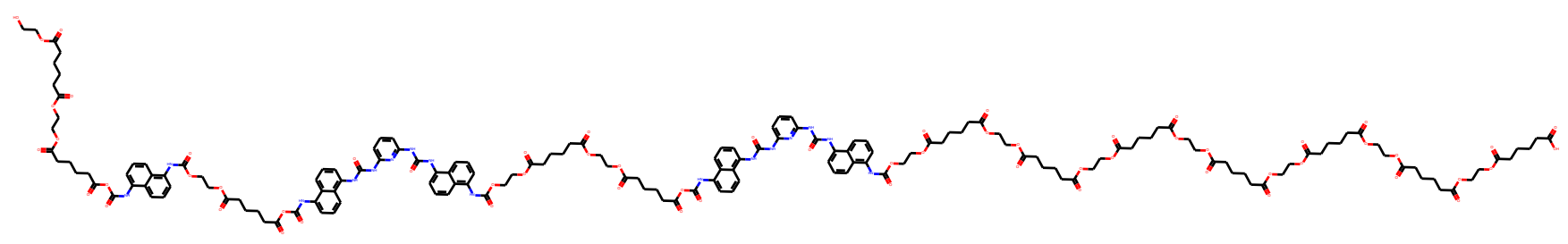 | | | |
| 4 | DBDI | PCL | MDA | $\mathcal{HHSHSHHSHHSHSHS}$ |
| | 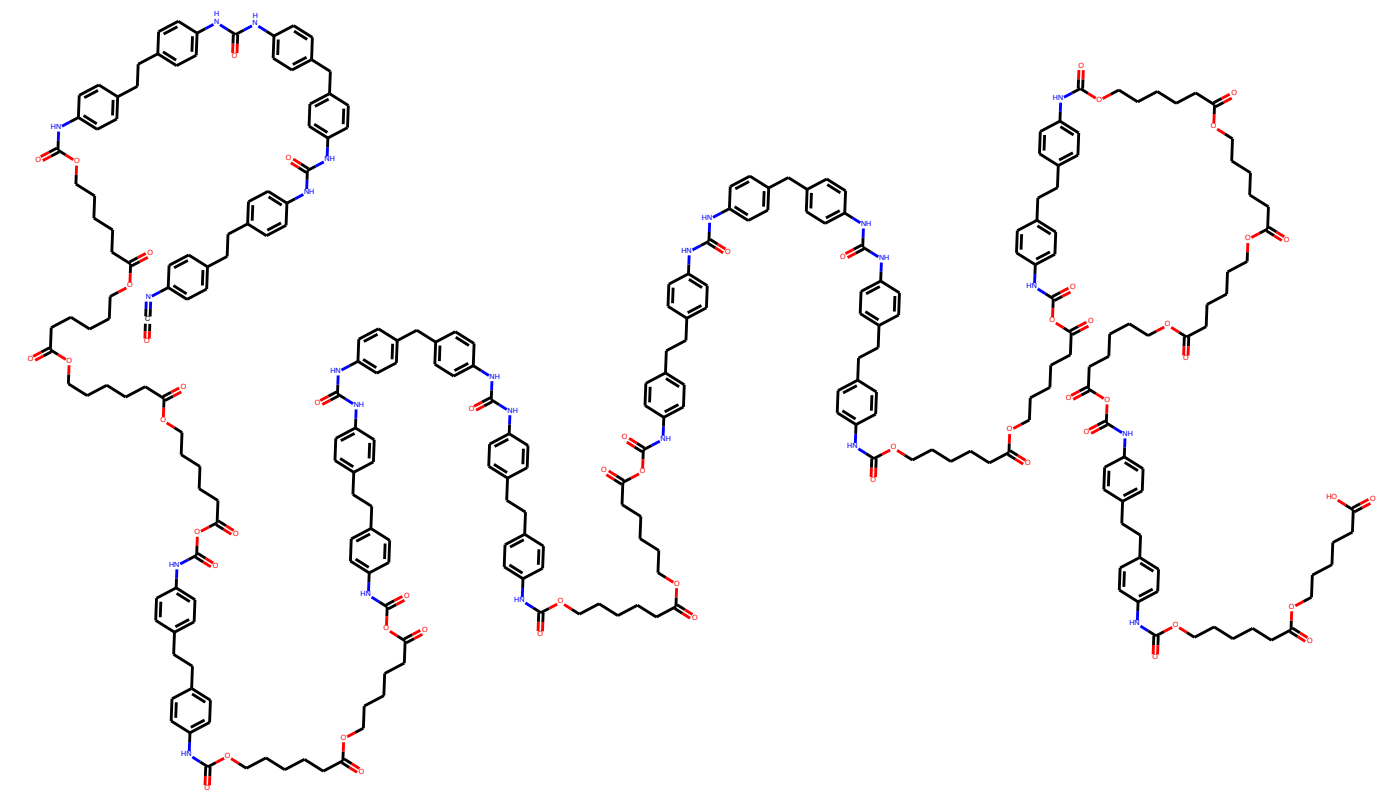 | | | |
| 5 | MDI | PHA | BDO | $\mathcal{SHSHHHSHS}$ |
| | 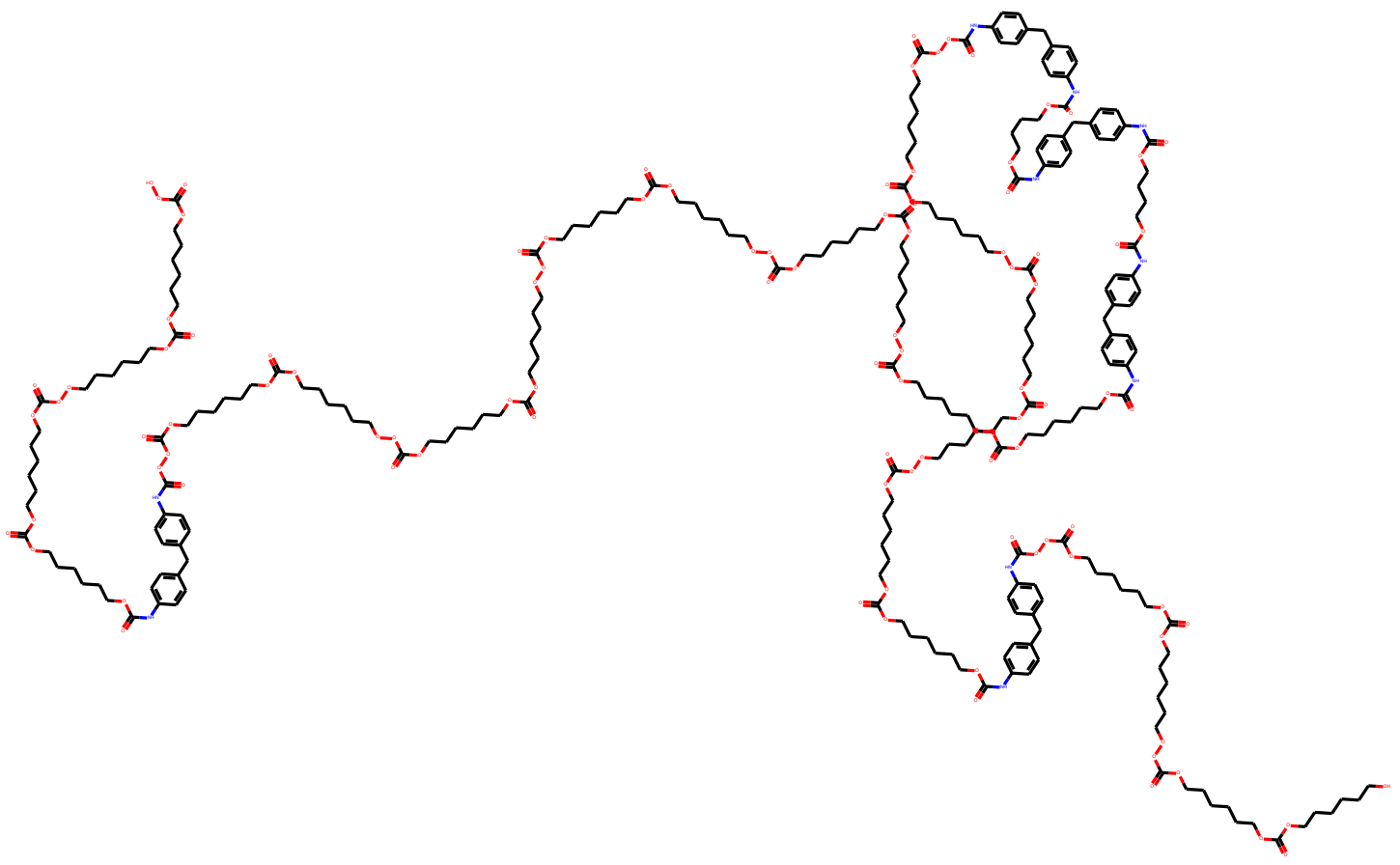 | | | |
| 6 | DBDI | Poly bd | DAB | $\mathcal{SHSHHHSHHHSHHSH}$ |
| | 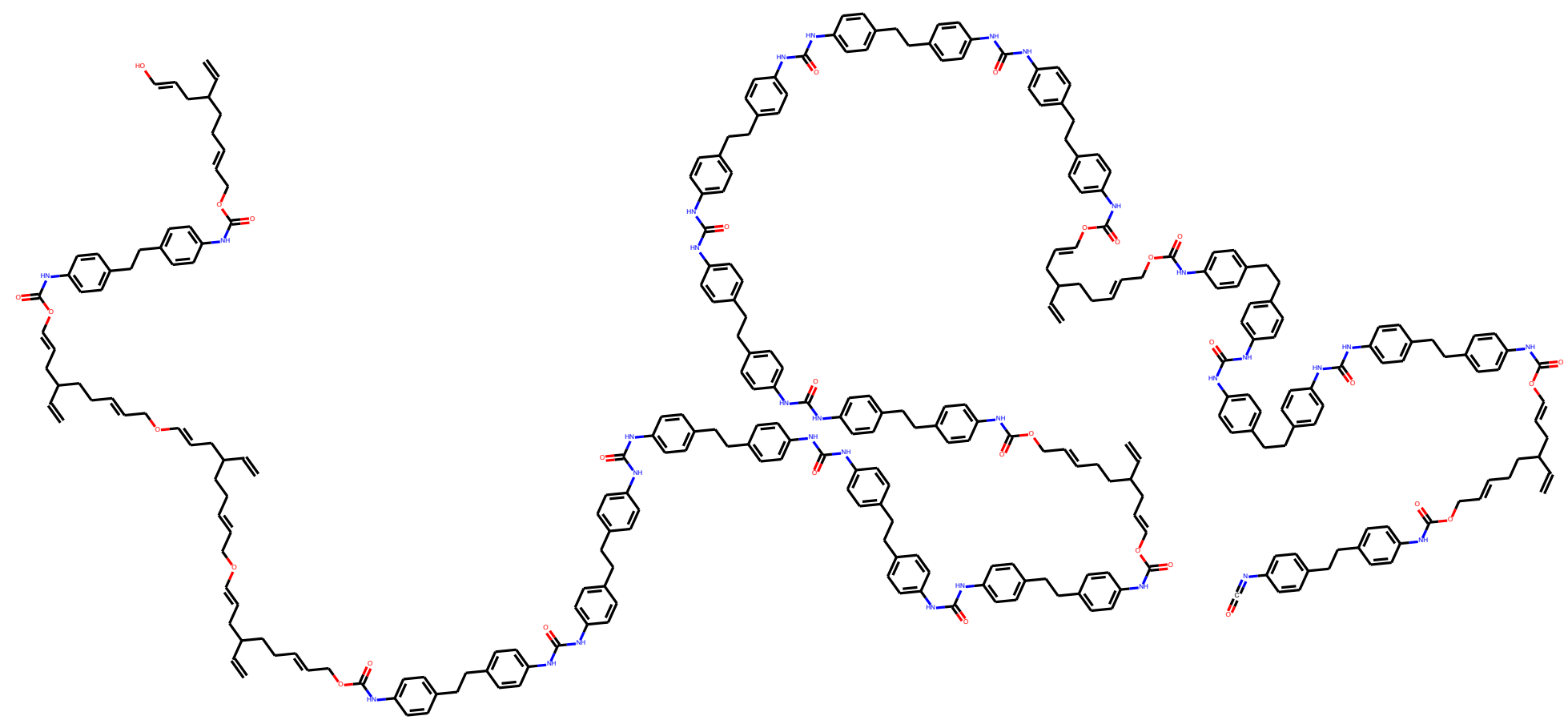 | | | |
| 7 | NDI | PTMO | DAB | $\mathcal{SHHHSHSHSHSHHS}$ |
| | 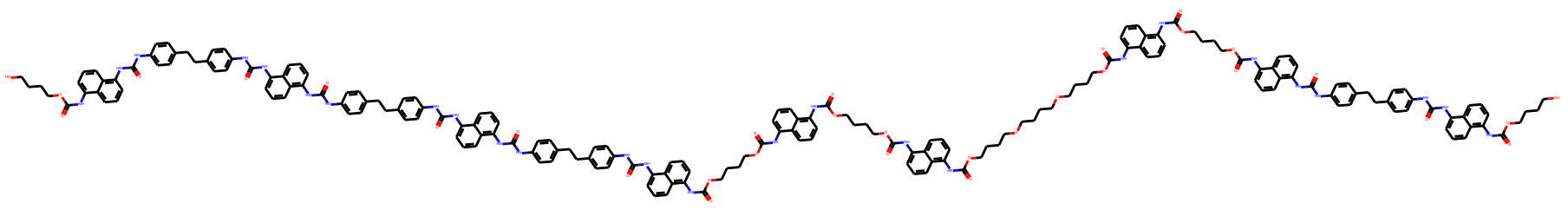 | | | |
| 8 | MDI | PTMO | EG | $\mathcal{SHHHSHSHSHSHS}$ |
| | 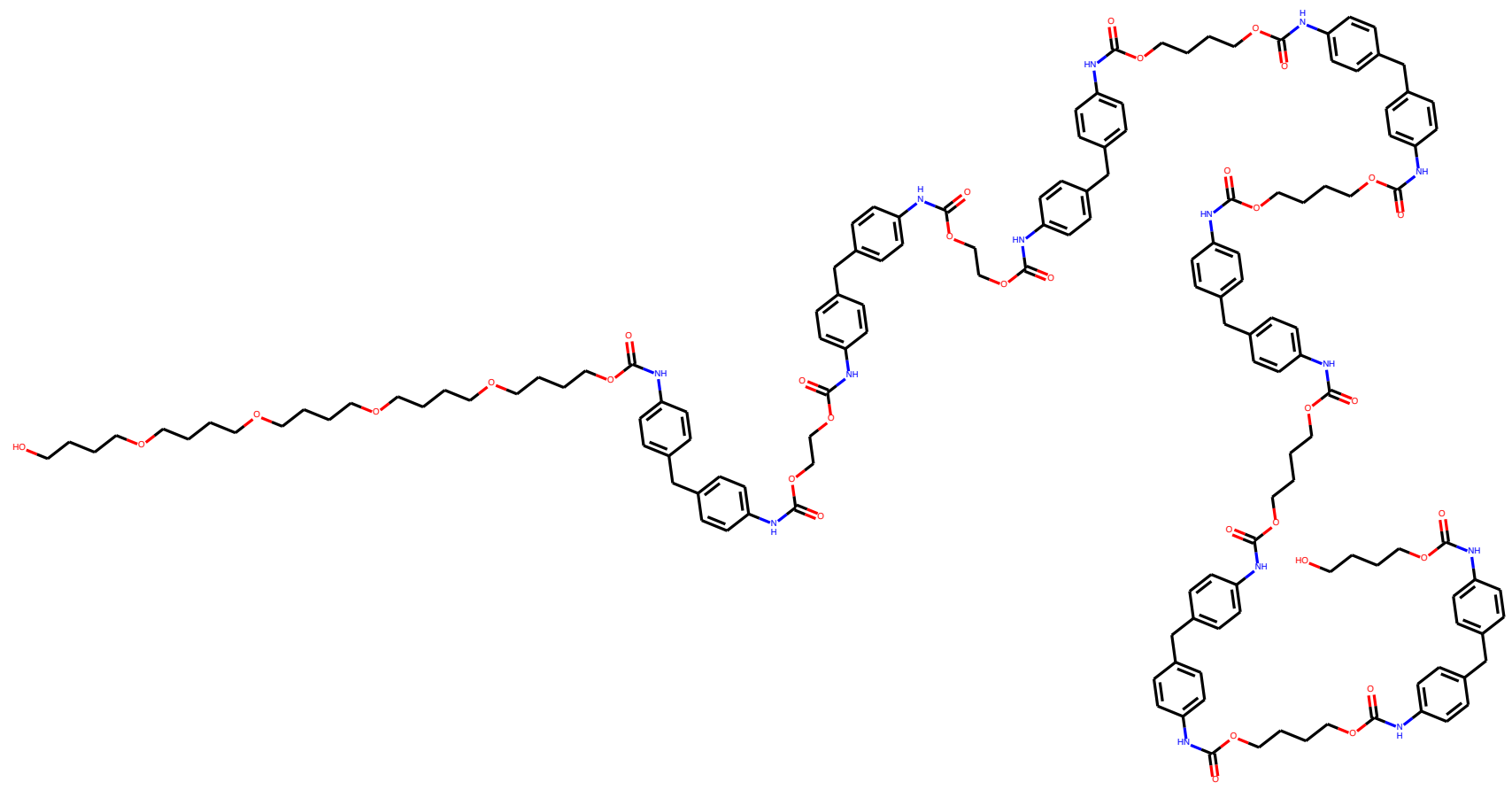 | | | |
| 9 | HDI | PBU | DAPy | $\mathcal{HSHSHHSHSHHHSHS}$ |
| | 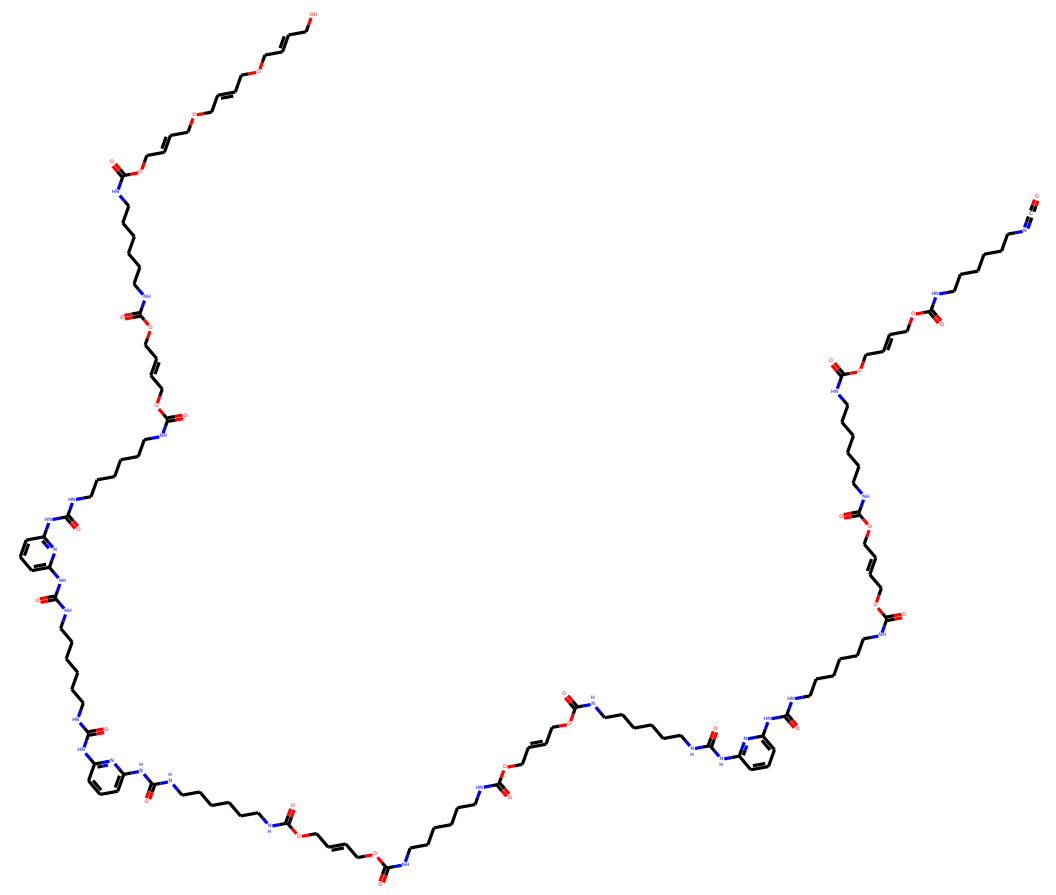 | | | |
| 10 | MDI | PCL | DAPy | $\mathcal{SHHHSHHSHHSHHHSH}$ |

| | |
|---|---|
| | 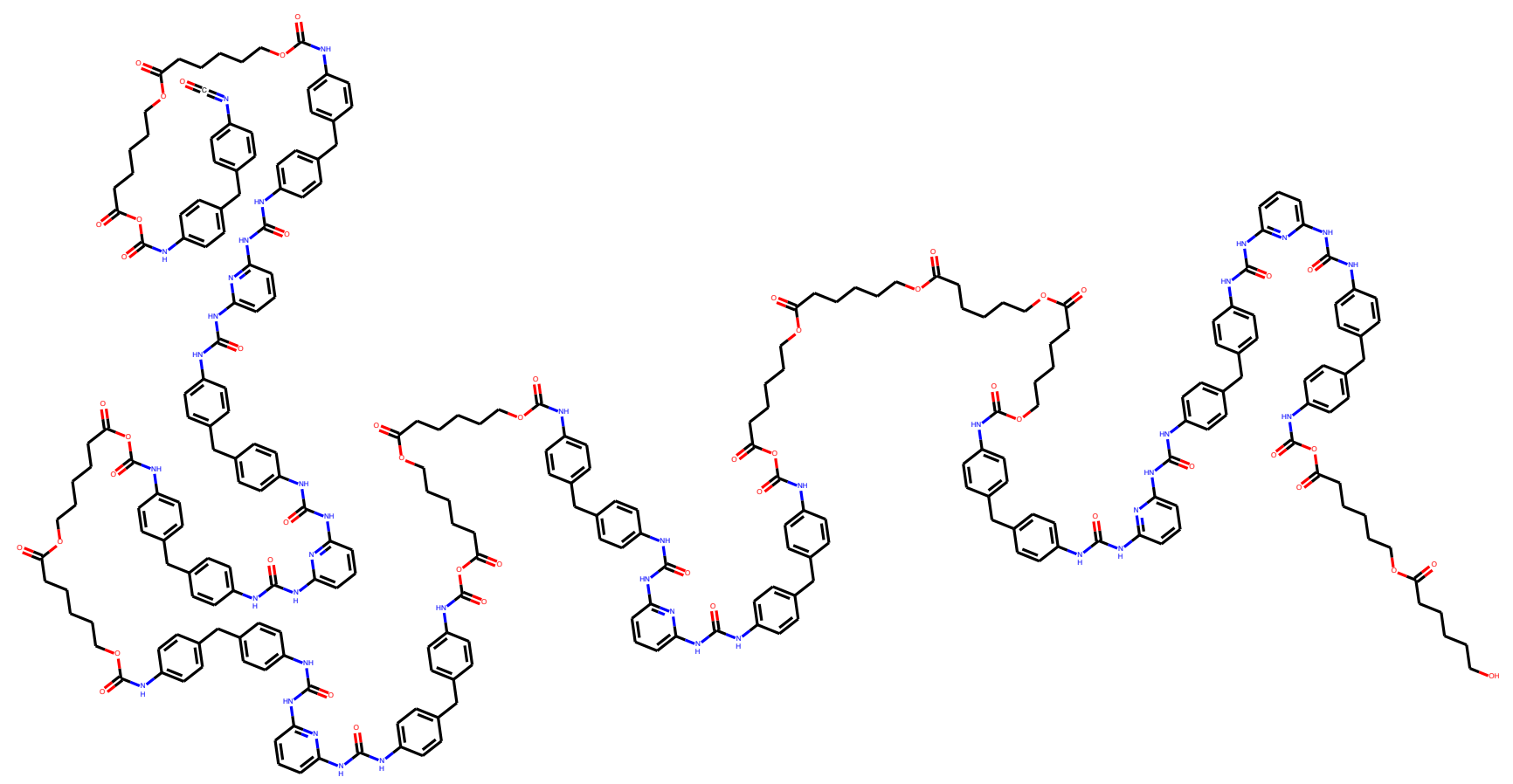 |

## S7. Examples of Branched Polyurethane Chains

| Index | Diisocyanate | Macrodiol | Chain Extender | Generated Hypergraph Symbolic String |
|---|---|---|---|---|
| 1 | TDI | PET | DAPy, 3THA | $\mathcal{H}\mathcal{S}\mathcal{H}\mathcal{H}\mathcal{H}\mathcal{H}\mathcal{H}[\mathcal{H}\mathcal{H}\mathcal{H}\mathcal{H}]\mathcal{S}\mathcal{H}\mathcal{H}\mathcal{H}$ |
| | 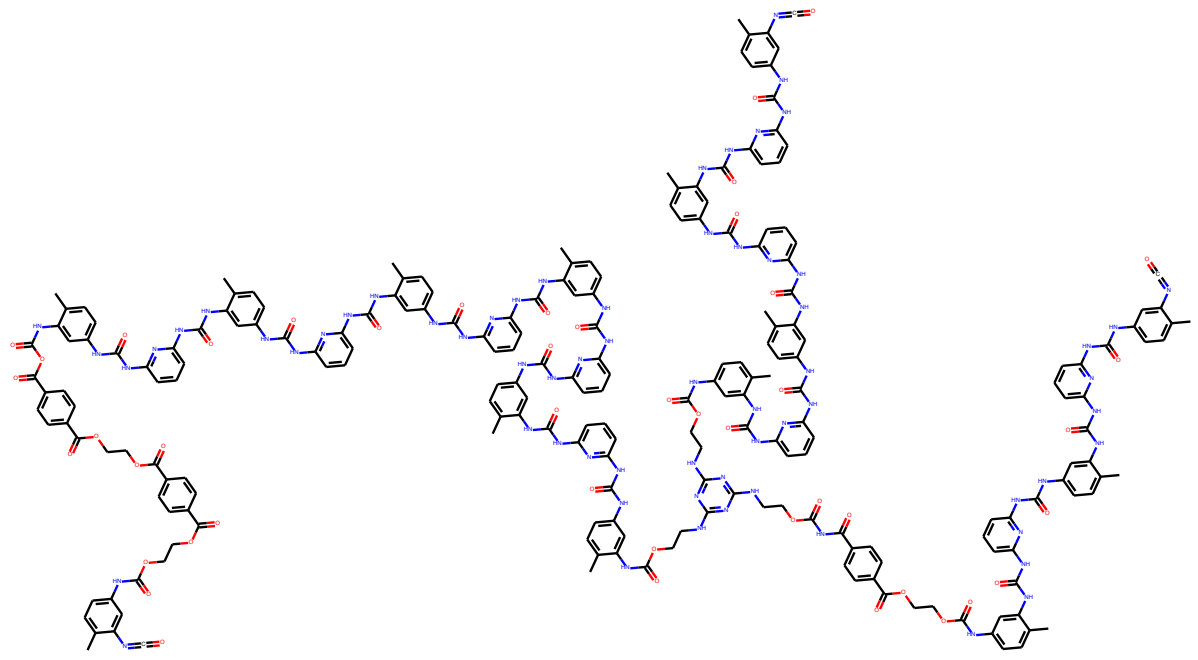 | | | |
| 2 | DBDI | PTMO | MDA, 3THA | $\mathcal{H}\mathcal{S}\mathcal{H}[\mathcal{S}\mathcal{H}]\mathcal{H}\mathcal{S}\mathcal{H}\mathcal{S}\mathcal{H}\mathcal{H}[\mathcal{H}[\mathcal{S}]\mathcal{H}[\mathcal{S}]\mathcal{H}]\mathcal{H}\mathcal{H}[\mathcal{S}]\mathcal{S}$ |
| | 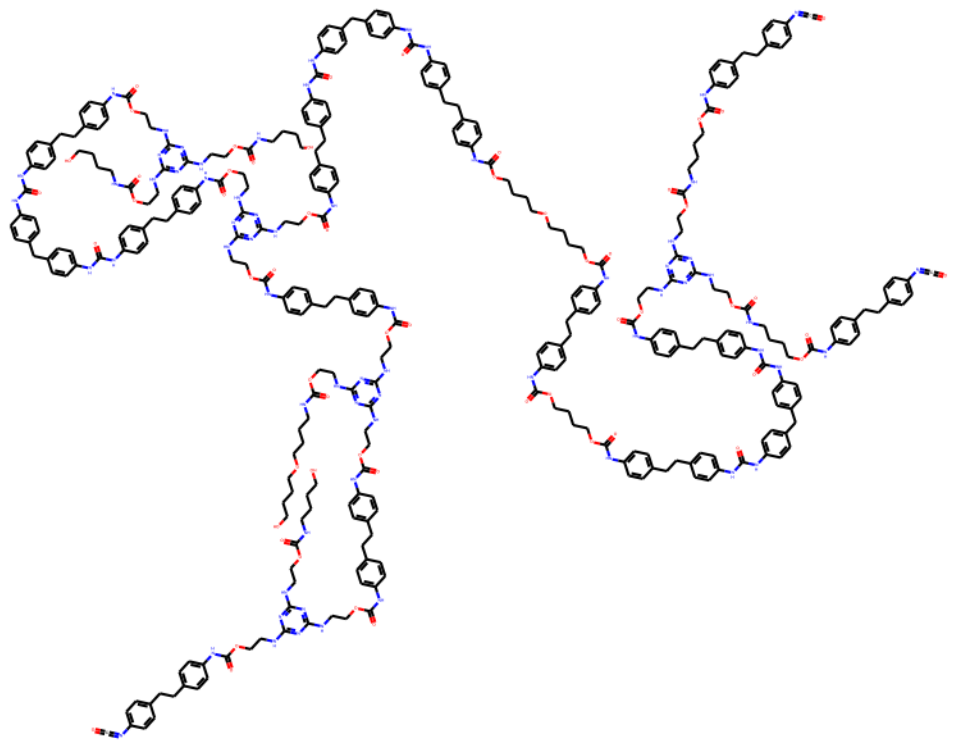 | | | |
| 3 | DBDI | PCD | EG, 3THA | $\mathcal{H}\mathcal{H}[\mathcal{H}]\mathcal{H}\mathcal{H}\mathcal{S}\mathcal{H}\mathcal{S}\mathcal{H}\mathcal{H}\mathcal{S}\mathcal{H}$ |
| | 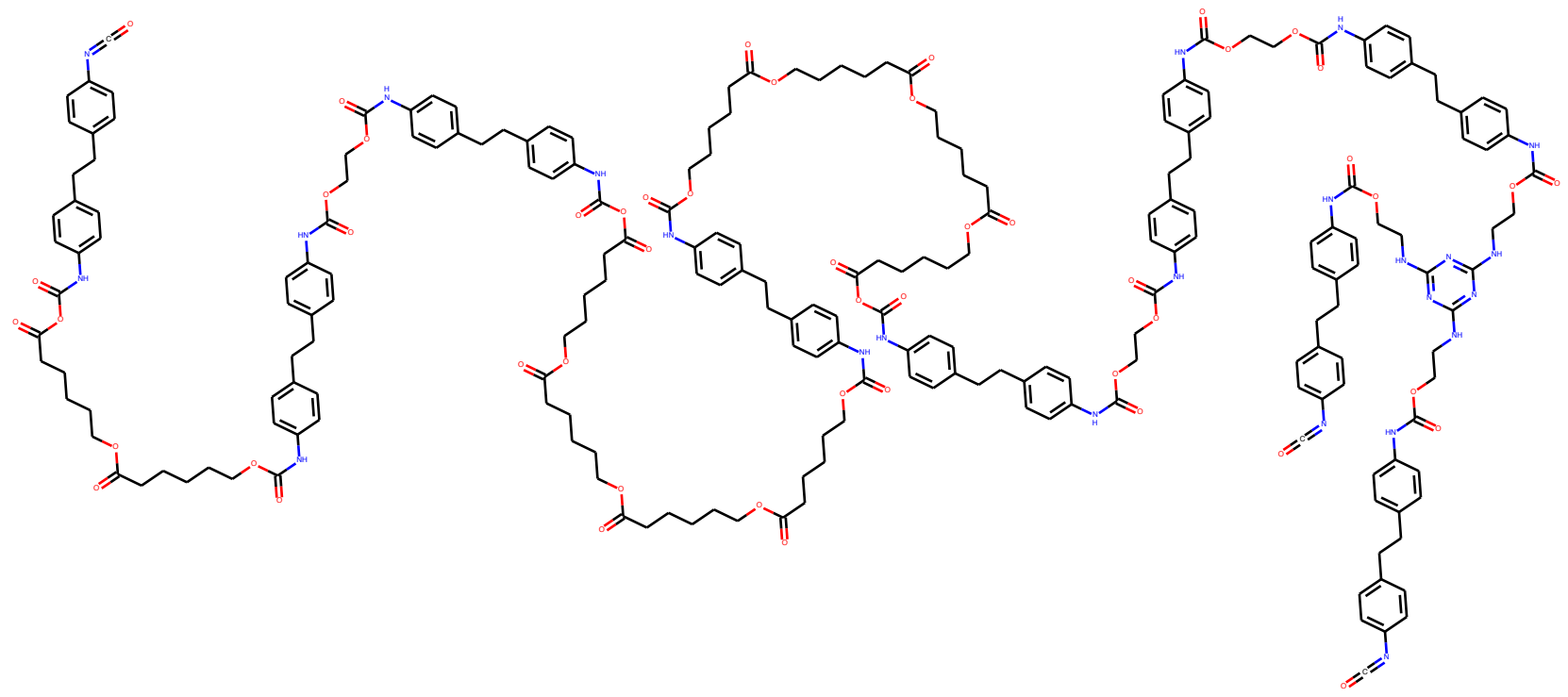 | | | |
| 4 | HDI | PBA | BG, 3THA | $\mathcal{HSHSHH}[\mathcal{HS}]\mathcal{SHHS}$ |
| | 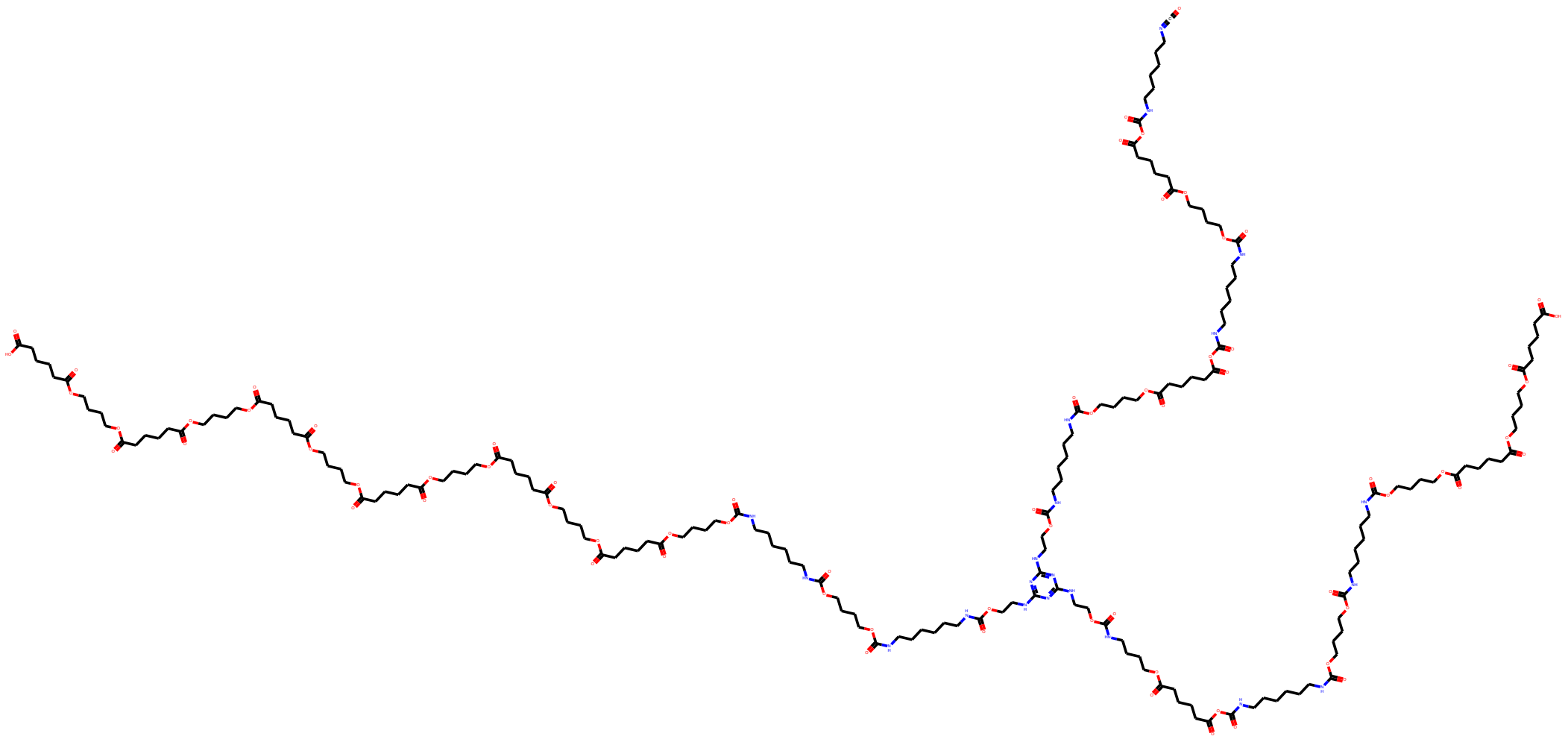 | | | |
| 5 | NDI | CHDM | EG, 3THA | $\mathcal{HSH}[\mathcal{S}]\mathcal{H}[\mathcal{HS}]\mathcal{HHS}$ |
| | 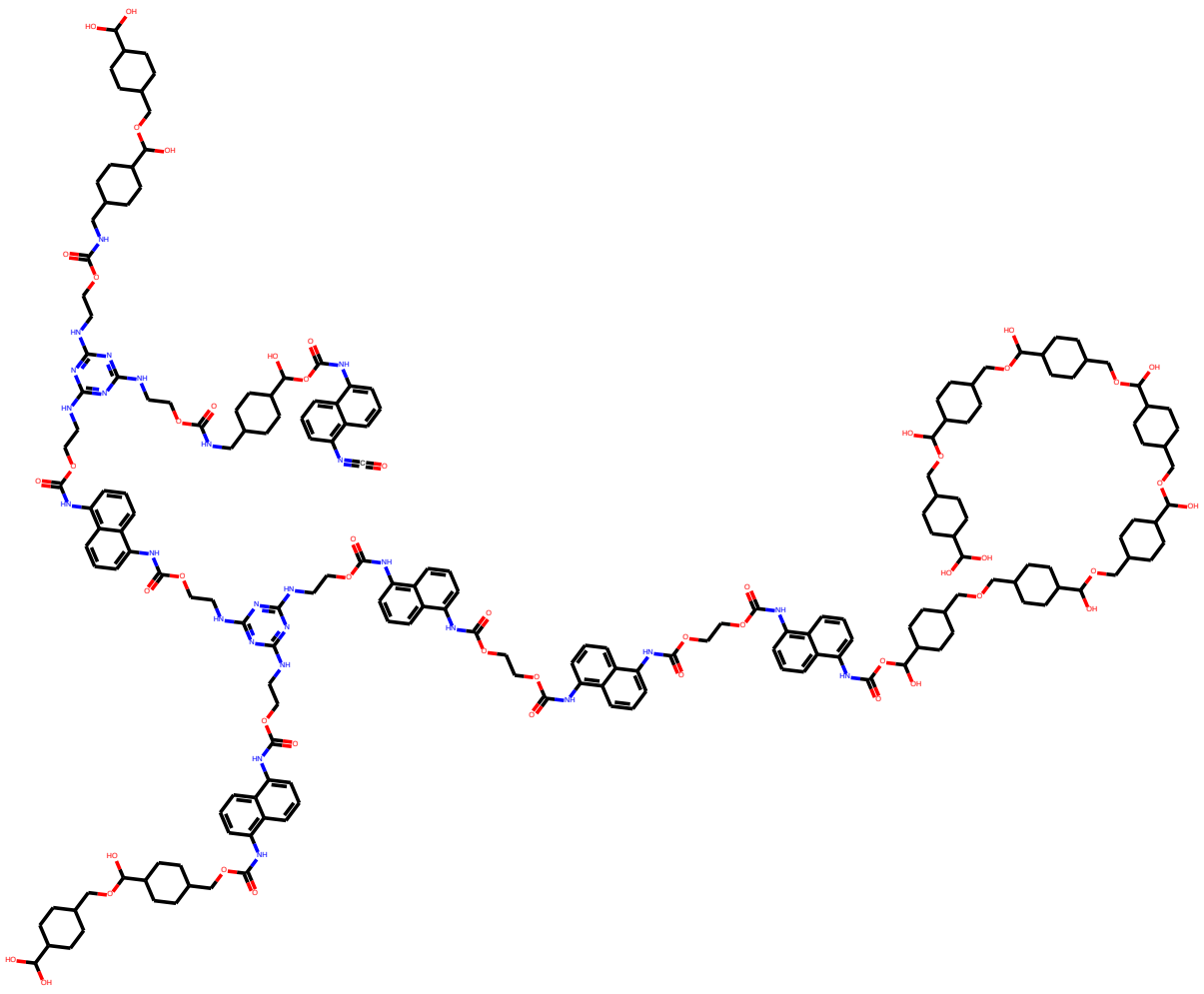 | | | |

## S8. Examples of Acrylate's Functional Groups

| Index | Generated Hypergraph Symbolic String | Acrylate's Functional Group |
|---|---|---|
| 1 | b(1)c(4)[b(1)c(1)][b(1)c(1)]b(1)c(4)[b(1)c(1)][b(1)c(1)]b(1)c(2) | 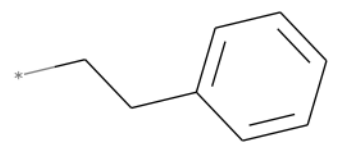 |
| 2 | b(1)c(4)[b(1)c(1)][b(1)c(1)]b(1)c(4)[b(1)c(1)][b(1)c(1)]b(1)c(4)[b(1)c(1)][b(1)c(1)]b(1)c(4)[b(1)c(1)][b(1)c(1)]b(1)c(4)[b(1)c(1)][b(1)c(1)]b(1)c(4)[b(1)c(1)][b(1)c(1)]b(1)c(4)[b(1)c(1)][b(1)c(1)]b(1)c(4)[b(1)c(1)][b(1)c(1)]b(1)c(1) | 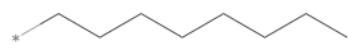 |
| 3 | b(1)c(4)[b(1)c(4)[b(1)c(1)][b(1)c(1)]b(1)c(1)][b(1)c(1)]b(1)c(4)[b(1)c(1)][b(1)c(1)]b(1)c(4)[b(1)c(1)][b(1)c(1)]b(1)c(1) | 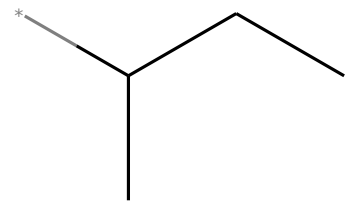 |
| 4 | b(1)c(4)[b(1)c(4)[b(1)c(1)][b(1)c(1)]b(1)c(1)][b(1)c(4)[b(1)c(1)][b(1)c(1)]b(1)c(1)][b(1)c(4)[b(1)c(1)][b(1)c(1)]b(1)c(1)] | 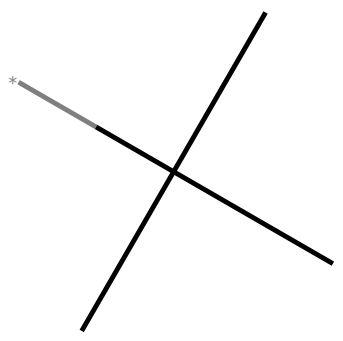 |
| 5 | b(1)c(4)[b(1)c(1)][b(1)c(1)]b(1)c(4)[b(1)c(1)][b(1)c(1)]b(1)c(2)b(1)c(4)b(1)c(1)[b(1)c(1)][b(1)c(1)] | 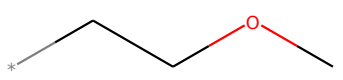 |
| 6 | b(1)c(2)b(1)c(4)[b(1)c(1)][b(1)c(1)]b(1)c(2)b(1)c(4)[b(1)c(1)][b(1)c(1)]b(1)c(1) | 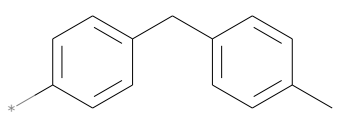 |
| 7 | b(1)c(4)[b(2)c(2)]b(1)c(3)[b(1)c(1)]b(1)c(1) | 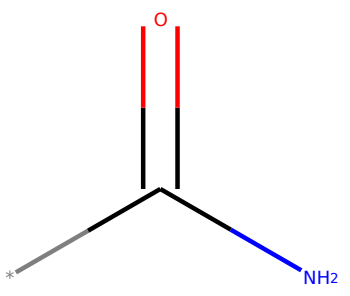 |
| 8 | b(1)c(2)b(2)c(2)b(1)c(4)[b(1)c(1)][b(1)c(4)[b(1)c(1)][b(1)c(1)]b(1)c(1)b(1)c(4)[b(1)c(1)][b(1)c(1)]b(1)c(1) | 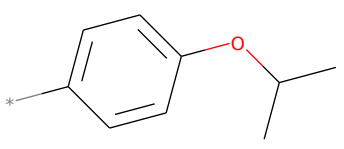 |
| 9 | b(1)c(4)[b(1)c(2)b(1)c(1)][b(1)c(1)]b(1)c(4)[b(1)c(4)[b(1)c(1)][b(1)c(1)]b(1)c(1)][b(1)c(4)[b(1)c(1)][b(1)c(1)]b(1)c(1)][b(1)c(4)[b(1)c(1)][b(1)c(1)]b(1)c(1)] | 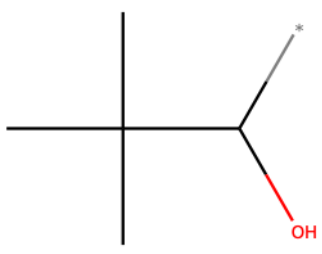 |
| 10 | $b(1)c(2)b(1)c(2)b(1)c(4)[b(1)c(4)[b(1)c(1)][b(1)c(1)]b(1)c(1)][b(1)c(4)[b(1)c(1)][b(1)c(1)]b(1)c(1)][b(1)c(4)[b(1)c(1)][b(1)c(1)]b(1)c(1)]b(1)c(2)b(1)c(2)b(1)c(1)$ | 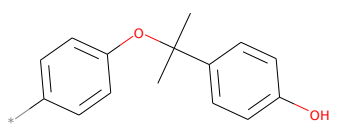 |

## S9. Abbreviations and Acronyms

| | |
|---|---|
| MDI | 4,4’-methylenebis(phenyl isocyanate) |
| TDI | toluene-diisocyanate |
| DBDI | 4,4’-dibenzyl diisocyanate |
| HDI | 1,6-diisocyanatohexane |
| HMDI | hydrogenated MDI |
| IPDI | isophorone diisocyanate |
| NDI | 1,5-Naphthalene diisocyanate |
| TMDI | 2,2,4-trimethyl-1,6-hexamethylelne diisocyanate |
| PEG / PEO | poly(oxyethylene) glycol |
| PEA | poly(ethylene adipate)diol |
| PBA | poly(butane adipate) diol |
| PTMO / PTHF | poly(oxytetramethylene) diol / polytetrahydrofurane diol |
| PBU | poly(butadiene)diol |
| PCL / PCD | polycaprolactone diol |
| PHA | polyhexamethylene carbonate glycol |
| PET | polyethylene terephthalate |
| PLA | polylactic acid (lactic acid) |
| CHDM | 1,4-cyclohexane dimethanol |
| Poly bd | polybutadiene diol |
| BD / BG / BDO | 1,4-butanediol |
| EG | ethylene glycol |
| DEG | diethylene glycol |
| DAPO | 2,5-bis-(4-amino-phenylene)-1,3,4-oxadiazole |
| DAB | 4,4’-diamino-dibenzyl |
| DAPy | 2,6-diamino-pyridine |
| MDA | 4,4'-methylene-dianiline |